\documentclass[sn-basic]{sn-jnl}

\jyear{2023}%

\usepackage{xcolor}
\usepackage{subcaption}
\usepackage{graphicx}
\usepackage{pdflscape}
\usepackage{afterpage}
\usepackage{capt-of}

\usepackage[shortlabels]{enumitem}

\newcommand{\btheta}{{\boldsymbol{\theta}}}

\usepackage{placeins} 

\begin{document}

\title[Quantifying model discrepancy using an ensemble of experimental designs]{Empirical quantification of predictive uncertainty due to model discrepancy by training with an ensemble of experimental designs: an application to ion channel kinetics}


\author[1]{\fnm{Joseph G.} \sur{Shuttleworth}}

\author[2,3]{\fnm{Chon Lok} \sur{Lei}}

\author[1]{\fnm{Dominic G.} \sur{Whittaker}}

\author[4,5]{\fnm{Monique J.} \sur{Windley}}

\author[4,5]{\fnm{Adam P.} \sur{Hill}}

\author[1]{\fnm{Simon P.} \sur{Preston}}

\author*[1]{\fnm{Gary R.} \sur{Mirams}}\email{gary.mirams@nottingham.ac.uk}

\affil[1]{\orgdiv{Centre for Mathematical Medicine \& Biology, School of Mathematical Sciences}, \orgname{University of Nottingham}, \orgaddress{\street{University Park}, \city{Nottingham}, \postcode{NG7 2RD}, \country{United Kingdom}}}

\affil[2]{\orgdiv{Institute of Translational Medicine, Faculty of Health Sciences}, \orgname{University of Macau}, \state{Macau}, \country{China}}

\affil[3]{\orgdiv{Department of Biomedical Sciences, Faculty of Health Sciences}, \orgname{University of Macau}, \state{Macau}, \country{China}}


\affil[4]{\orgdiv{Computational Cardiology Laboratory},
\orgname{Victor Chang Cardiac Research Institute},
\city{Darlinghurst}, \state{New South Wales}, \country{Australia}}

\affil[5]{\orgdiv{School of Clinical Medicine, Facility of Medicine and Health,}, \orgname{University of New South Wales},
\city{Sydney}, \state{New South Wales}, \country{Australia}}

\abstract{
When using mathematical models to make quantitative predictions for clinical or industrial use, it is important that predictions come with a reliable estimate of their accuracy (uncertainty quantification).
Because models of complex biological systems are always large simplifications, model discrepancy arises---models fail to perfectly recapitulate the true data generating process. 
This presents a particular challenge for making accurate predictions, and especially for accurately quantifying uncertainty in these predictions.

Experimentalists and modellers must choose which experimental procedures (\emph{protocols}) are used to produce data used to train models. 
We propose to characterise uncertainty owing to model discrepancy with an ensemble of parameter sets, each of which results from training to data from a different protocol. 
The variability in predictions from this ensemble provides an empirical estimate of predictive uncertainty owing to model discrepancy, even for unseen protocols. 

We use the example of electrophysiology experiments that investigate the properties of hERG potassium channels. 
Here, `information-rich' protocols allow mathematical models to be trained using numerous short experiments performed on the same cell. 
In this case, we simulate data with one model and fit it with a different (discrepant) one. 
For any individual experimental protocol, parameter estimates vary little under repeated samples from the assumed additive independent Gaussian noise model. 
Yet parameter sets arising from the same model applied to different experiments conflict--- highlighting the model discrepancy.
Our methods will help select more suitable ion channel models for future studies, and will be widely applicable to a range of biological modelling problems.
}




\maketitle

\newpage
\section{Introduction}\label{sec:introduction}

Mathematical models are used in many areas of study to provide accurate quantitative predictions of biological phenomena. 
When models are used in safety-critical settings (such as drug safety or clinical decision-making), it is often important that our models produce accurate predictions over a range of scenarios, for example, for different drugs and patients. Perhaps more importantly, these models must allow a reliable quantification of confidence in their predictions.
The field of \emph{uncertainty quantification} (UQ) is dedicated to providing and communicating appropriate confidence in model predictions \citep{smith2013uncertainty}.
Exact models of biological phenomena are generally unavailable, and we resort to using approximate mathematical models instead. 
When our mathematical model does not fully recapitulate the data-generating process (DGP) of a real biological system, we call this \emph{model discrepancy} or model misspecification. This discrepancy between the DGP and our models presents a particular challenge for UQ.

Often, models are trained using experimental data from a particular experimental design, and then used to make predictions under (perhaps drastically) different scenarios.
We call the set of experimental designs under consideration the \emph{design space} and denote it \(\mathcal{D}\).
We assume the existence of some DGP, which maps each element of \(d \in \mathcal{D}\) to some random output. 
These elements are known as experimental designs, or, as is more common in electrophysiology, \emph{protocols}, and each corresponds to some scenario that our model can be used to make predictions for. 
Namely, in Section~\ref{sec:simple_example}, each protocol, \(d\in\mathcal{D}\), is simply a set of observation times.
By performing a set of experiments (each corresponding to a different protocol \(d\in\mathcal{D}\)) we can investigate (and quantify) the difference between the DGP and our models in different situations.
When training our mathematical models using standard frequentist or Bayesian approaches, it is typically assumed that there is no model discrepancy; in other words, that the data arise from the model (for some unknown, true parameter set). 
This is a necessary condition for some desirable properties of the parameter estimators which provide some guarantees regarding the accuracy of parameter estimates when there is a large number of observations, as discussed in Section~\ref{sec:discrepancy}. 
However, when model discrepancy is not considered, we can find that the ability of a model to make accurate predictions is compromised. In particular, if we try to validate our model with a protocol dissimilar to that used for training, there can be a noticeable difference between our predictions and the data---even when the model appears to fit the training data well.
A simple illustration of this problem is introduced in the following section. 

\subsection{Motivating Example} \label{sec:simple_example}
In this section, we construct a simple example where we train a discrepant model with data generated from a DGP using multiple experimental designs. 
This example demonstrates that it is important to consider the protocol-dependence of parameter estimates and predictions when using discrepant models.

First, we construct a DGP formed of the sum of two exponential terms,
\begin{align}
    y^*\!(t) &= \exp\left\{-t\right\} + \exp\left\{-\frac{t}{10}\right\}, \label{eqn:toy_deterministic} \\\
    z^*\!(t) &= y^*\!(t) + \varepsilon(t),\label{eqn:toy_observations}
\end{align} 
for some \(t>0\) where \(\varepsilon(t)\) is an independent Gaussian random variable,
each with zero mean and variance, \(\sigma^2 = 10^{-4}\) for each \(t>0\). 
Here, \( z^*(t)\) is a random variable representing an observation of the system at some time, \(t\). 

Next, we attempt to fit a model which takes the form of single exponential decay,
\begin{align}
    y(t; \btheta) &= \theta_1 \exp\left\{-\frac{t_i}{\theta_2}\right\},\\ \label{eqn:simple_example_discrepant_y}
    z(t; \btheta) &= y(t; \btheta) + \varepsilon(t_i),
\end{align} 
to these data, denoting the column matrix \([\theta_1, \theta_2]^T\)  by \(\btheta\). 
We call this a discrepant model because there is no choice of \(\btheta\) such that \(y(t; \btheta) = y^*\!(t)\), for all \(t > 0 \).

To train our model, we choose a set of \(n\) observation times, \(T = \{t_1,\, t_2,\, \ldots\, t_n\}\).  
We may then find the parameter set, \(\hat\btheta(T)\), which minimises the sum-of-squares error between our discrepant model (Equation~\ref{eqn:simple_example_discrepant_y}) and each \(z(\btheta; t_i)\), that is,
\begin{equation}
    \hat\btheta(T) = \text{argmin}_{\btheta\in\Theta}\left\{ \sum_{t_i\in T} \big(y(t_i; \btheta) - z^*\!(t_i)\big)^2\right\}, \label{eqn:toy_estimation}
\end{equation}
where $T$ is a set of observation times.

\begin{figure}[htb]
    \centering
    \includegraphics{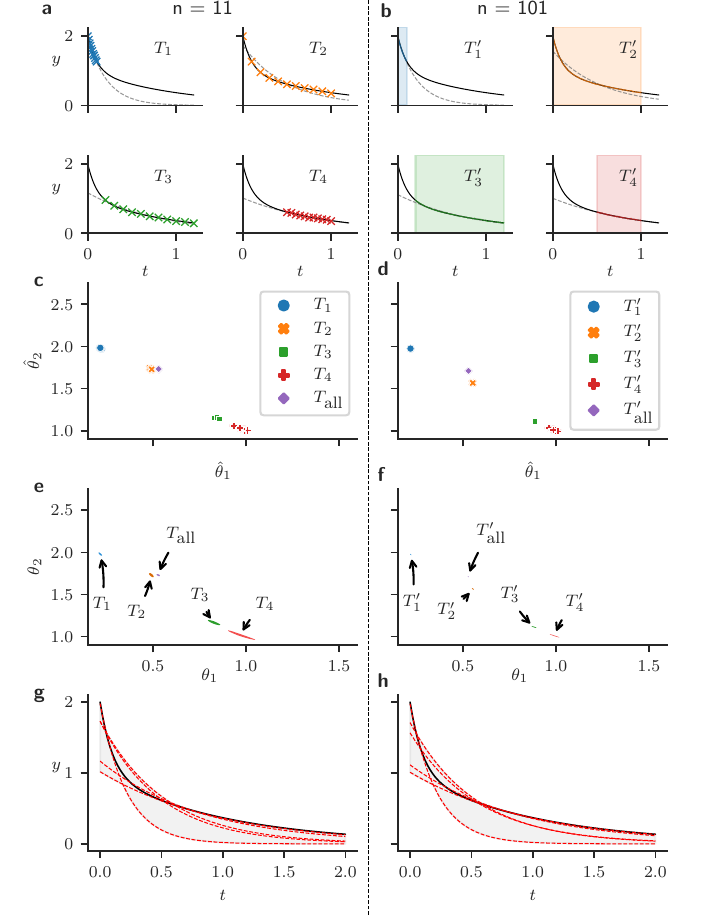}
    \caption{Under model discrepancy, parameter estimates depend on the design used for training. 
    Panels to the left of the dotted line correspond to designs containing \(n=11\) observations at times (\(T_1, \ldots, T_4\) as shown in panel \textbf{a}). 
    Panels on the right show designs with \(n=101\) observations, (\(T_1', \ldots, T_4'\) as shown in panel \textbf{b})
    (\textbf{a}) and (\textbf{b}): representative datasets generated by the DGP shown with the solid black line (Equation~\ref{eqn:toy_deterministic}) with points indicating observations (sampled using Equation~\ref{eqn:toy_observations}) and the fitted discrepant model (Equation~\ref{eqn:simple_example_discrepant_y}), with calibrated \(\btheta\)) (grey dashed lines). 
    (\textbf{c}) and (\textbf{d}): the parameter estimates for each design, each fitted to one of ten repeats of the DGP.
    (\textbf{e}) and (\textbf{f}): 99\% Bayesian credible regions obtained using MCMC, a uniform prior and a single repeat of the DGP.
    (\textbf{g}) and (\textbf{h}): predictions using the discrepant model fitted using a single repeat of each protocol (using the estimates shown in \textbf{e} and \textbf{f}), showing the true DGP (black), discrepant model predictions (red), and the difference between predictions (grey).}
    \label{fig:simple_example}
\end{figure}

Then, we consider multiple experimental protocols which we may use to fit this  model (Equation~(\ref{eqn:simple_example_discrepant_y})). 
In particular, we consider the following sets of observation times
%
\begin{align}
    T_1 &= \left\{0,\, 0.01,\, 0.02,\,  \ldots, 0.01\right\}, \\
    T_2 &= \left\{0,\, 0.1,\, 0.2,\, 0.3,\, \ldots,\, 1\right\}, \\
    T_3 &= \left\{0.2,\, 0.3,\, 0.4,\, 0.5,\, \ldots,\, 1.2\right\}, \\ 
    T_4 &= \left\{0.5,\, 0.55,\, 0.6,\, \ldots,\, 1 \right\}, \;\text{ and } \\ 
    T_\text{all} &= T_1 \cup T_2 \cup T_3 \cup T_4.
\end{align} 
We sample from the DGP \(10\) times by computing Equation~\ref{eqn:toy_observations} for each observation time, \(t\), and adding IID Gaussian noise. Then, for each sample of the DGP, we compute parameter estimates using each set of observation times (\(T_1, T_2, T_3, T_4\) and \(T_\text{all}\)). This process is then repeated with a ten-fold increase in sampling rate, that is, with observation times,
\begin{align}
    T_1' &= \left\{0,\, 0.001,\, 0.002,\,  \ldots, 0.01\right\}, \\
    T_2' &= \left\{0,\, 0.01,\, 0.02,\, 0.03,\, \ldots,\, 1\right\}, \\
    T_3' &= \left\{0.2,\, 0.21,\, 0.22,\, 0.23,\, \ldots,\, 1.2\right\}, \\ 
    T_4' &= \left\{0.5,\, 0.505,\, 0.51,\, \ldots,\, 1 \right\}, \;\text{ and }\\ 
    T_\text{all}' &= T_1 \cup T_2 \cup T_3 \cup T_4.
\end{align}

If we choose a Bayesian approach to the problem, we may specify a (relatively uninformative) uniform prior distribution on the model parameters, that is,
\begin{align}
    \theta_1 &\sim U(0, 10), \\ 
    \text{and } \; \theta_2 &\sim U(0, 10). 
\end{align} 
The likelihood of our misspecified model is
\begin{equation}
    \mathcal{L}(\btheta; \mathbf{z}^*) = \prod_{i=1}^n \sqrt{ \dfrac{1}{2\pi\sigma^2} } \exp\left\{\dfrac{\left( z^*(t_i)- y(t_i; \btheta) \right)^2 }{2 \sigma^2} \right\}, \label{eqn:simple_example_likelihood}
\end{equation} 
where \(n\) is the number of observations for this protocol, and \(\mathbf{z}^* = (z(t_i))_{i=1}^n\) is a vector of observations of the DGP. 
We may then explore the posterior distribution \citep{gelman2013bayesian} using Markov chain Monte Carlo. 
In particular we use the Haario-Bardenet adaptive-covariance Metropolis-Hastings \citep{johnstone_uncertainty_2016} algorithm as implemented by PINTS \citep{clerx_probabilistic_2019}. 
This method was run using four parallel chains each containing 25,000 iterations and a `burn in' period of 5,000 iterations. 

As shown in Fig.~\ref{fig:simple_example}, for each sample of the DGP, we obtain a parameter estimate from each set of observation times, each with a different distribution. 
For instance, training using \(T_1\) results in a model that approximates the DGP well on short timescales, and training using \(T_4\) allows us to recapitulate the behaviour of the system over longer timescales, as can be seen in panel \textbf{a}. 
From how closely the discrepant model (Equation~\ref{eqn:simple_example_discrepant_y}) fits the data in the regions where observations are made (in Fig.~\ref{fig:simple_example}, panels \textbf{a} and \textbf{b}), we can see that in either case, a single exponential seems to provide a reasonable approximation to the DGP. 
However, if we require an accurate model for both the slow and fast behaviour of the system, model discrepancy presents an issue, and this model (namely, Equation~\ref{eqn:simple_example_discrepant_y}) may be unsuitable. This is the case for \(T_2\) as shown in Fig.~\ref{fig:simple_example}\textbf{a}. This variability in behaviour is shown in Fig.~\ref{fig:simple_example}, panels \textbf{g} and \textbf{h}, which show how the model's predictions for \(0 \leq t \leq 2\) depend on what protocol was used to fit the model. 

The Bayesian posteriors illustrated in Fig.~\ref{fig:simple_example}, panels \textbf{e} and \textbf{f}, show that we are not able to avoid the problems caused by model discrepancy by simply adopting a Bayesian framework---we will obtain precise parameter estimates that are highly dependent on the chosen training protocol, nevertheless. This problem becomes more obvious when we increase the number of observations. In the examples detailed in this paper, we explore this `high-data limit' where the variability in each parameter estimate (under repeated samples of the DGP) is minuscule compared to the difference between parameter estimates obtained from different protocols.

\subsection{Ion channel modelling}

Discrepancy has proven to be a difficult problem to address in modelling electrically-excitable cells \citep[\emph{electrophysiology} modelling, ][]{lei_considering_2020, mirams2016uncertainty}. 
The same is true for many other mathematical models, such as rainfall-runoff models in hydrology \citep{beven_manifesto_2006}, models of the spread of infectious diseases in epidemiology \citep{guan_modeling_2020,creswell2023epidemic}, and models used for the prediction of financial markets \citep{ANDERSON2009233}.

The `rapid delayed rectifier potassium current' (I\textsubscript{Kr}), carried by the channel encoded primarily by hERG, plays an important role in the recovery of heart cells in from electrical stimulation. 
It allows the cell membrane to return to its `resting potential' ahead of the next heartbeat. 
This current can be blocked by pharmaceutical drugs, disrupting this process and causing dangerous changes to heart rhythm. 
Mathematical models  are now routinely used in drug safety assessment to test whether the expected dynamics of I\textsubscript{Kr} under drug block are expected to cause such problems \citep{li_improving_2017}.
However, these models provide only an incomplete description of I\textsubscript{Kr}, and do not, for example, account for the stochastic behaviour of individual ion channels \citep{mirams2016uncertainty}. 
For this reason, an understanding of model discrepancy and its implications is crucial in building accurate dynamical models of I\textsubscript{Kr} which permit a realistic appraisal of their predictive uncertainty.

In Section~\ref{sec:simple_example}, we presented a simple example, in which each protocol corresponds to a particular choice of observation times. However, there may other aspects of the design to be decided upon. For example, in electrophysiology, whole-cell patch-clamp experiments are performed by placing an electrode in the solution inside the cell membrane (the intracellular solution), and another in the solution outside the cell (the extracellular solution). 
Voltage-clamp experiments are a particular type of patch-clamp experiment in which a voltage signal is applied across the cell membrane, whilst the current flowing across the cell membrane is recorded. Here, the protocol consists of the chosen voltage for each time (treated as a forcing function in ODE-based models), together with a set of observation times for the resulting current (observed output).

Electrophysiologists have a lot of control, and therefore choice, regarding the protocol design; but little work has been done to explore how the choice of protocol used to gather training data affects the accuracy of subsequent predictions. 
We explore these protocol-dependent effects of model discrepancy in Section~\ref{sec:results}. 

\subsection{Previously proposed methods to handle discrepancy}

One way of reducing over-confidence in inaccurate parameter estimates in the presence of model discrepancy may be to use approximate Bayesian computation (ABC) \citep{https://doi.org/10.1111/rssb.12356}. 
With ABC, a likelihood function is not explicitly specified; instead, the model is repeatedly simulated for proposed values of the parameter sampled from a prior distribution. 
Each proposed value is accepted or rejected according to whether the simulated trajectory is ``close'' to the actual data, according to some chosen summary statistics. 
ABC compares the simulated with the real data using these summary statistics (rather than matching all aspects of the dynamics) and accepts approximate matches (subject to a chosen tolerance). It is suited to inference where there is substantial model discrepancy because this approach can decrease potential over-confidence in the inferred values of parameters. 
However, it is challenging to select suitable summary statistics, and the computational demands of ABC are much greater than those of the methods we propose.

Another approach was first introduced by \citet{kennedy_bayesian_2001}, who introduced Gaussian processes to the observables. 
This work has since been applied to electrophysiology models \citep{lei_considering_2020}.
Elsewhere, \citeauthor{sung_calibration_2020} introduced an approach to account for heteroscedastic errors using many repeats of the same experiment \citep{sung_calibration_2020}, although this seems to be less applicable to the hERG modelling problem (introduced in Section~\ref{sec:herg_modelling}) because the number of repeats of each experiment (when training individual, cell-specific models) is limited.
Alternatively, \citet{lei_neuralode_2021} modelled the discrepancy using a neural network within the differential equations. 
However, these approaches reduce the interpretability of otherwise simple mechanistic models, and, when compared with models that simply ignore model discrepancy, could potentially result in worse predictions under protocols that are dissimilar to those used for training.

Instead, we use a diverse range of experiments to train our models and build a picture of how  model discrepancy manifests under different  training protocols. 
We are then able to judge the suitability of our models, and provide empirically-derived, \emph{spread-of-prediction} intervals which provide a realistic level of predictive uncertainty due to model discrepancy. 
We demonstrate the utility of these methods under synthetically generated data by constructing two examples of model discrepancy.




\section{Methods} \label{sec:methods}
We begin with a general overview of our proposed methods before providing two real-world examples of their applications.
In Section~\ref{sec:partially_observable}, we outline some notation for a statistical model consisting of a dynamical system, an observation  function, and some form of observational noise. 
This allows us to talk, in general terms, about model calibration and validation in Section~\ref{sec:fitting}. 
In particular, we describe a method for validating our models, in which we change the protocol used to train the model. 
This motivates our proposed methods for combining parameter estimates obtained from different protocols to empirically quantify model discrepancy for the prediction of unseen protocols.

\subsection{Fitting models using multiple experimental protocols}\label{sec:general_method}

\subsubsection{Partially observable ODE models}\label{sec:partially_observable}
In this paper, we restrict attention to deterministic models of biological phenomena, in which a system of \emph{ordinary differential equations} (ODEs) is used to describe the deterministic time-evolution of some finite number of states. Although, the method would generalise to other types of models straightforwardly. 
This behaviour may be dependent on the protocol, \(d\), chosen for the experiment, and so, we express our ODE system as,
\begin{align}
    \dfrac{\textrm{d}\mathbf{x}}{\textrm{d}t} &= \mathbf{f}(\mathbf{x}, t; \btheta_f, d), \label{eqn:general_ode}
\end{align}
 where \(\mathbf{x}\) is a column vector of length \(N\) describing the `state' of the system, \(t\) is time, and the parameters specifying the dynamics of the system are denoted \(\btheta_f\). 
 Additionally, the system is subject to some initial conditions which may be dependent on \(\btheta_f\). 
 Owing to \(\mathbf{x}\)'s dependence on the protocol and model parameters, we use the notation,
\begin{equation}
    \mathbf{x}(t; \btheta_f, d),
\end{equation} 
to denote the solution of Equation~\ref{eqn:general_ode} under protocol \(d\) and a specific choice of parameters, \(\btheta_f\).

This ODE system is related to our noise-free observables via some \emph{observation function} of the form,
\begin{equation}
      h\left(\mathbf{x}, t; \btheta_{h}, d\right),
\end{equation} 
where \(\mathbf{x}\) is the state of the ODE system (Equation~\ref{eqn:general_ode}), \(t\) is the time that the observation occurs, \(d\) is the protocol, and some additional parameters \(\btheta_h\), which are distinct from those in \(\btheta_f\). 
Here, we make observations of the system, via this function, at a set of observation times, \(\{t_i\}_{i=1}^{n_d}\) defined by the protocol, \(d\). 

For concision, we may stack \(\btheta_f\) and \(\btheta_{h}\) into a single vector of model parameters,
\begin{equation}
    \btheta = \begin{bmatrix} \btheta_f \\
    \btheta_h
    \end{bmatrix}.
\end{equation}
Then, we denote an observation at time \(t_i\)
by 
\begin{equation}
    y_i(\btheta; d) = h\big(\mathbf{x}(t_i; \btheta_f, d), t_i; \btheta_{h}, d\big), \label{eqn:y_defn}
\end{equation} 
We denote the set of possible model parameters by \(\Theta\), such that 
$\btheta \in \Theta$.
We call this collection of possible parameter sets the \emph{parameter space}.

For each protocol, \(d\in\mathcal{D}\), and vector of model parameters, \(\btheta\), we may combine each of our observations into a vector, 
\begin{equation}
\mathbf y(\btheta; d) = 
\begin{bmatrix}
    y_1(\btheta; d), \\
    \vdots \\
    y_{n_d}(\btheta; d)
\end{bmatrix}.
\end{equation}

Additionally, we assume some form of random observational error such that, for each protocol, \(d\), each observation is a random variable,
\begin{equation}
    z_i(d) = y_i(\btheta; d) + \varepsilon_i, \label{eqn:general_observation}
\end{equation} 
where each \(\varepsilon_i\) is the error in the \(i^\text{th}\) observation. 
Here  each protocol, \(d\), is performed exactly once so that we obtain one sample of each vector of observations (\(\mathbf{z}(d)\)). 
In the examples presented in Sections \ref{sec:CaseI} and \ref{sec:caseII}, we assume that our observations are subject to independent and identically distributed (IID) Gaussian errors, with mean, \(0\), and standard deviation, \(\sigma\). 

\subsubsection{Evaluation of predictive accuracy and model training} \label{sec:fitting}
Given some parameter set \(\btheta\), we may evaluate the accuracy of the resultant predictions under the application of some protocol \(d\in\mathcal{D}\) by computing the root-mean-square error (RMSE) \citep{https://doi.org/10.1029/JC090iC05p08995} between these predictions, and our observations (\(\mathbf z(d)\)), 
\begin{equation}
    \text{RMSE}\big(\mathbf{y}(\btheta; d),\;\mathbf{z}(d)\big) = \sqrt{\dfrac{1}{n_{d}}\sum_{i=1}^{n_{d}} \big(y_i(\btheta; d) - z_i(d)\big)^2},
\label{eqn:rmse}
\end{equation} 
where \(n_d\) is the number of observations in protocol \(d\). 
We choose the RMSE as it permits comparison between protocols with different numbers of observations.  

Similarly, we may train our models to data, \(\mathbf{z}(d)\), obtained using some protocol, \(d\), by finding the parameter set that minimises this quantity  (Equation~\ref{eqn:rmse}). 
In this way, we define the parameter estimate obtained from protocol \(d\) as,
\begin{equation}
    \hat\btheta_d = \text{argmin}_{\btheta\in\Theta}\left\{\text{RMSE}\big(\mathbf{y}(\btheta, d),\; \mathbf{z}(d)\big)\right\}, \label{eqn:parameter_estimation}
\end{equation} 
which is a random variable (because it depends on our random data, \(\mathbf{z}\)).
Since minimising the RMSE is equivalent to minimising the sum-of-squares error, this estimate is also the least-squares estimator (identical to Equation~\ref{eqn:toy_estimation}). 
Moreover, under the assumption of Gaussian IID errors, Equation~\ref{eqn:parameter_estimation} is exactly the \emph{maximum likelihood estimator} because the natural logarithm of the likelihood can be written as, 
\begin{equation}
    \log\left\{ \mathcal{L}(\btheta; \mathbf{z})\right\} = -\dfrac{n}{2} \log \left(2\pi\hat \sigma^2\right) - \dfrac{1}{2\sigma^2} \sum_{i=1}^n \big(y_i(\btheta; d) - z_i(d)\big)^2, \label{eqn:log_likelihood}
\end{equation} where \(\hat \sigma\) is an estimate of \(\sigma\). 
Equation~\ref{eqn:log_likelihood} can be minimised by first finding the parameter set, \(\hat \btheta\) which minimises the sum-of-squares error term, then finding the optimal \(\sigma\). 
 Whilst these estimates of \(\btheta\) are identical whether or not \(\sigma\) is known, only examples with known (and not estimated) \(\sigma\) are explored in this paper.

Having obtained such a parameter estimate, we may validate our model, by computing predictions for some other protocol, \(\tilde d\in\mathcal{D}\). 
To do this, we compute, 
\(
    \mathbf{y}(\hat\btheta_d; \tilde d).
\)
This is a simulation of the behaviour of the system (without noise) under protocol \(\tilde d\) made using parameter estimates that were obtained by training the model to protocol \(d\) (as in Equation~\ref{eqn:parameter_estimation}). 
In this way, our parameter estimates, each obtained from different protocols, result in different out-of-sample predictions (predictions for the results for protocols other than the one used for training). 
Because we aim to train a model able to produce accurate predictions for all \(d\in\mathcal{D}\), it is important to validate our model using multiple protocols. 

By computing \(\text{RMSE}\big(\mathbf{y}(\hat\btheta_d; \tilde d), \mathbf z(\tilde d)\big)\) for each pair of training and validation protocols, \(d\) and  \(\tilde d\),  we are able to perform model validation across multiple training and validation protocols. 
This allows us to ensure our models are robust with regard to the training protocol, and allow for the quantification of model discrepancy as demonstrated in Section~\ref{sec:results}.

\subsubsection{Consequences of model error/discrepancy} \label{sec:discrepancy}

Ideally, we would have a model that is correctly specified, in the sense that the data arise from the model being fitted. 
In other words, our observations \(\mathbf{z}(d)\) arise from Equation~\ref{eqn:general_observation} where \(\btheta\) is some fixed, unknown value, \(\btheta^*\in\Theta\). 
Then we may consider the distance between an estimate \(\hat\btheta\) and the true value. 
When the model is correctly specified and given suitable regularity conditions on the model and the design, $d$, we can obtain arbitrarily accurate parameter estimates by increasing the number of observations, \(n\). That is, more precisely, that \(\hat\btheta\) converges in probability to \(\btheta^*\) as \(n \rightarrow \infty\).  This property is known as consistency \citep{seber_nonlinear_2005}.
These regularity conditions include that the model is structurally identifiable for the particular \(d\in\mathcal{D}\) used for training, for example. That is, different values of the parameter, \(\btheta\), result in different model output \citep{wieland_structural_2021}. Other conditions ensure that \(\mathbf{y}(\btheta; d)\) is suitably smooth as a function of \(\btheta\) \citep{seber_nonlinear_2005}. 

However, when training discrepant models, we may find that our parameter estimates are heavily dependent on the training protocol, as demonstrated in Section~\ref{sec:simple_example}. 
For unseen protocols, these discordant parameter sets may lead to a range of vastly different predictions, even if each parameter set provides a reasonable fit for its respective training protocol. 
In such a case, further data collection may reduce the variance of these parameter estimates, but fail to significantly improve the predictive accuracy of our models.  

In Section~\ref{sec:results}, we explore two examples of synthetically constructed model discrepancy. 
In Section~\ref{sec:CaseI}, we have that \(\mathbf f\) and \(h\) (Equations~\ref{eqn:general_ode} and \ref{eqn:general_observation}) are exactly those functions used to generate the data, and the exact probability distribution of the observational errors is known. However, one parameter is fixed to an incorrect value. 
In other words, the true parameter set \(\btheta^*\) lies outside the parameter space used in training the model. 
Under the assumption of structural identifiability (and a compact parameter space), this is an example of model discrepancy because there is some limit to how well our model can recapitulate the DGP. 

In Section~\ref{sec:caseII} we explore another example of model discrepancy where our choice of \(\mathbf{f}\) (and, in this case, the dimensions of \(\btheta\) and \(\mathbf{x}\)) are misspecified by training a model which differs structurally from the one used in the DGP.


\subsubsection{Ensemble training and prediction interval}

As outlined in Section~\ref{sec:fitting}, we can obtain parameter estimates from each protocol \(d\in\mathcal{D}\) by finding the \(\hat\btheta\in\Theta\) that minimises Equation~\ref{eqn:parameter_estimation}. 
We then obtain an \emph{ensemble} of parameter estimates,
\begin{equation}
    \left\{\hat\btheta_d \; : \; d \in \mathcal{D}_\text{train}\right\}. \label{eqn:set_of_estimates}
\end{equation}
Then, for any validation protocol \(\tilde d\), the set, 
\begin{equation}
    \left\{ \mathbf y (\hat\btheta_d; \tilde d) \; : \; d \in \mathcal{D}_{\text{train}}\right\}, \label{eqn:set_of_predictions}
\end{equation} 
is an ensemble of predictions where \(\mathcal{D}_\text{train}\subseteq\mathcal{D}\) is some set of training protocols. 
Each of these estimates may be used individually to make predictions. 
We may then use these ensembles of parameter estimates to attempt to quantify uncertainty in our prediction. 
We do this by considering the range of our predictions for each observation of interest. 
For the \(i\text{th}\) observation of our validation protocol, \(\tilde d\), that is
\begin{align}
\mathcal{B}^{(i)}
&= \left[\mathcal{B}_\text{lower}^{(i)},\mathcal{B}_\text{upper}^{(i)}\right] \nonumber \\
&= \left[\min_{d\in\mathcal{D_\text{train}}}\left\{
    y_i(\hat\btheta_d; \tilde d)
    \right\},
    \max_{d\in\mathcal{D_\text{train}}}\left\{
    y_i(\hat\btheta_d; \tilde d)
    \right\}
    \right], \label{eqn:spread_of_prediction_interval}
\end{align} 

When all observations are considered at once, Equation~\ref{eqn:spread_of_prediction_interval} comprises a band of predictions, giving some indication of uncertainty in the predictions. 
We demonstrate below that this band provides a useful indication of predictive error for unseen protocols, and provides a range of plausible predictions. We propose that a wide band of predictions for a given validation protocol suggests that there is model discrepancy and poor prediction accuracy for a particular context of use.

This interval (Equation~\ref{eqn:spread_of_prediction_interval}), cannot shrink as more protocols are added. If a large number of protocols are considered, percentiles of our ensemble of predictions may provide additional insight. However, in this paper, we only consider cases where there are a small number of protocols (five training protocols are used in each of the examples discussed in Section~\ref{sec:results}).

For the purposes of a point estimate, we use the midpoint of each interval,
\begin{equation}
    \mathcal{B}^{(i)}_\text{mid} = \dfrac{\mathcal{B}^{(i)}_\text{lower} + \mathcal{B}^{(i)}_\text{upper}}2. \label{eqn:midpoint_prediction}
\end{equation}  This is used to assess the predictive error of the ensemble in Fig.~\ref{fig:example2_error_compare}. 
There are other ways to gauge the central tendency of the set of predictions (Equation~\ref{eqn:set_of_predictions}). 
Such a change would have little effect on Section~\ref{sec:results}, but a median or weighted mean may be as (or more) suitable for other problems.

\subsection{Application to an ion current model} \label{sec:herg_modelling}


We now turn our attention to an applied problem in which dynamical systems are used to model cellular electrophysiology. 
We apply our methods to two special cases of model discrepancy using synthetically generated data.

Firstly, we introduce a common paradigm for modelling macroscopic currents in electrically excitable cells, so-called \emph{Markov models} \citep{rudy_computational_2006,fink_markov_2009}. 
In this setting, the term `Markov model' is often used to refer to systems of ODEs where the state variables describe the proportional occupancy of some small collection of `states', and the model parameters affect transition rates between these states. 
These models are discussed in Section~\ref{sec:markov_models} and may be seen as a special case of the more general ODE model introduced in Section~\ref{sec:general_method}. Additionally, in Section~\ref{sec:protocols}, we briefly introduce some relevant electrophysiology and in Section~\ref{sec:computational_methods}, we provide a detailed overview of our computational methods.

\subsubsection{Markov models of I\textsubscript{Kr}} \label{sec:markov_models}

Here, we use Markov models to describe the dynamics of I\textsubscript{Kr}, especially in response to changes in the transmembrane potential. 
For any Markov model (as described above), the derivative function can be expressed in terms of a matrix, \(\mathbf{A}\), which is dependent only on the transmembrane potential, \(V\). 
Accordingly, where \(x_i\) denotes the proportion of channels in some state, \(i\), Equation~\ref{eqn:general_ode} becomes,
\begin{align}
    \dfrac{\textrm{d}\mathbf{x}}{\textrm{d}t} &= \mathbf{f}(\mathbf x, t; \btheta_f, d), \nonumber \\
    &= \mathbf{A}\left(V(t; d); \btheta_f\right) \mathbf x, \label{eqn:markov_ode}
\end{align} 
where \(V(t; d)\) is the specified transmembrane potential at the time \(t\) under protocol \(d\). 
The elements of \(\mathbf{A}(V; \btheta_f)\), that is, \(\mathbf{A}_{i, j}(V; \btheta_f)\) describe the \emph{transition rate} from state \(j\) to state \(i\) with transmembrane potential, \(V\). 
Usually, the transition rates (elements of \(\mathbf{A}\)) are either constant or of the form \(\theta_i e^{\pm \theta_j V(t; d)}\) with \(\theta_i, \theta_j > 0\). Hence, each transition rate, \(k\) is either \(0\) for all \(V \in \mathbb{R}\) or satisfies \(k > 0\) for all \(V \in \mathbb{R}\).

Before and after each protocol, cells are left to equilibrate with the voltage \(V\) set to the \emph{holding potential}, \(V_\text{hold}=-80\)\,mV. 
Therefore, we require the initial conditions, for at time \(t=0\), 
\begin{equation}
    \mathbf{x}(0) = \mathbf{x}_\infty(V_\text{hold}),
\end{equation} 
where, \(\mathbf{x}_\infty(V_\text{hold})\) is the unique steady-state solution for the linear system, 
\begin{equation}
\dfrac{\textrm{d}\mathbf{x}}{\text{d}t} = \mathbf{A}(V_\text{hold}; \btheta_f) \mathbf x,
\end{equation} 
subject to the constraint \(\sum_{i=1}^N x_i(0) = 1\). 
Note also that \(\mathbf{A}(V_\text{hold}; \btheta_f)\) may be singular, as is the case when the number of channels is conserved (\(\sum_{i=1}^Nx_i(t) = 1\) for all \(t\)). 
This is the case for both Markov models used in this paper. 
To find \(\mathbf{x}_\infty\), we may follow \citeauthor{fink_markov_2009}'s method  \citep[see Supplementary Material of][]{fink_markov_2009}. A more technical discussion of the steady states of such ODE systems is found in \citet{keizer1972solutions}.

As is standard for models of I\textsubscript{Kr} \citep{beattie2015mathematical}, we take our observation function to be 
\begin{equation}
 I_\text{Kr} =   h(\mathbf{x}, t_i; \theta_{h}, d) = g \cdot [O](t; \btheta_{f}, d) \cdot (V(t; d) - E_\text{Kr}), \label{eqn:markov_observation_func}
\end{equation} 
where \([O]\) denotes the proportion of channels in an `open' conformation (one of the components of \(\mathbf{x}\)); and \(g\) is the sole parameter in \(\btheta_h\), known as the \emph{maximal conductance}; and \(E_\text{Kr}\) is the Nernst potential.
\(E_\text{Kr}\) is found by calculating
\begin{equation}
    E_\text{Kr} = \dfrac{RT}{F}\ln\left\{\dfrac{[K_\text{out}]}{[K_\text{in}]}\right\}, \label{eqn:reversal}
\end{equation} 
where \(R\) is the gas constant, \(F\) is Faraday's constant, and \(T\) is the temperature and \([K_\text{in}]\) and \([K_\text{out}]\) are the intracellular and extracellular concentrations of \(K^+\), respectively. 
Here, we choose the temperature to be room temperature (\(T=298\text{K}\)), and our intracellular and extracellular concentrations to be 120\,mM and 5\,mM, respectively, which approximately correspond to physiological concentrations \citep{hille}. 
Hence, for all synthetic data generation, training, and validation in Sections~\ref{sec:CaseI} and \ref{sec:caseII}, we fix \(E_\text{Kr} = -80.24\)\,mV (using Equation~\ref{eqn:reversal}). 

From Equations~\ref{eqn:markov_ode}--\ref{eqn:markov_observation_func}, we can see that both the dynamics of the model and the observation function are dependent on the voltage, \(V(t; d)\).  
This is a special case of Equations~\ref{eqn:y_defn}--\ref{eqn:general_observation}, in which \(\mathbf{f}\) and \(h\) are time-dependent only via the voltage $V(t;d)$. 
In other words, at any fixed voltage (\(V\)), Equation~\ref{eqn:markov_ode} is an autonomous system and \(h\) does not depend directly on \(t\).

We assume that our observational errors are additive Gaussian IID random variables with zero mean and variance, \(\sigma^2\).

The first model of I\textsubscript{Kr} we consider is by \citet{beattie_sinusoidal_2018}. 
This is a four state Markov model with nine parameters (8 of which relate to the model's kinetics and form $\btheta_f$). 
We use the parameters that \citeauthor{beattie_sinusoidal_2018} obtained from one particular cell (cell \#5) by training their model to data obtained from an application of the `sinusoidal protocol' with a manual patch-clamp setup. 
The cells used were Chinese hamster ovary cells, which were made to heterologously over-express \emph{hERG1a}. 
These experiments were performed at room temperature. 

The second model is by \citet{wang_quantitative_1997}.
This is a five-state model which has \(15\) parameters.
These parameters were obtained by performing multiple voltage-clamp protocols, all at room temperature, on multiple frog oocytes overexpressing \emph{hERG}. 
These experiments are used to infer activation and deactivation time constants as well as steady-state current voltage relations, which are, in turn, used to produce estimates for the model parameters. 
Of the two parameter sets provided in \citet{wang_quantitative_1997}, we use the parameter set obtained by using the extracellular solution with a 2\,mM concentration of potassium chloride, as this most closely replicates physiological conditions.

The systems of ODEs for the Beattie and Wang models, as well as the parameterisations of the transition rates, are presented in Appendix~\ref{sec:appendix_A}. 
The values of the model parameters, as used in Sections~\ref{sec:CaseI} and \ref{sec:caseII} are given in Table~\ref{tab:parameter_sets}.


\begin{figure}[htb!]
    \centering
    \includegraphics{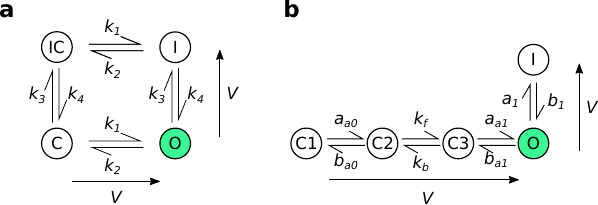}
    \caption{The structural differences between the two Markov model structures used in this paper for synthetic data generation and model training. 
    (\textbf{a}): the four-state Beattie model used in both \textit{Case I} and \textit{Case II}. 
    (\textbf{b}): the five-state Wang model used only for \textit{Case II}. 
    When a channel is in the open/conducting (\textit{O}) state (green) current is able to flow. 
    Whereas, when the model is in the other \emph{closed} (C) or \emph{inactivated} (I) states, no current can flow. 
    The arrows adjacent to each model structure indicate the direction in which rates increase as the voltage increases.
    }
    \label{fig:beattie_wang_diagrams}
\end{figure}




\subsubsection{Experimental Designs for Voltage Clamp Experiments} \label{sec:protocols}

\begin{figure}[htbp!]
    \centering
    \includegraphics{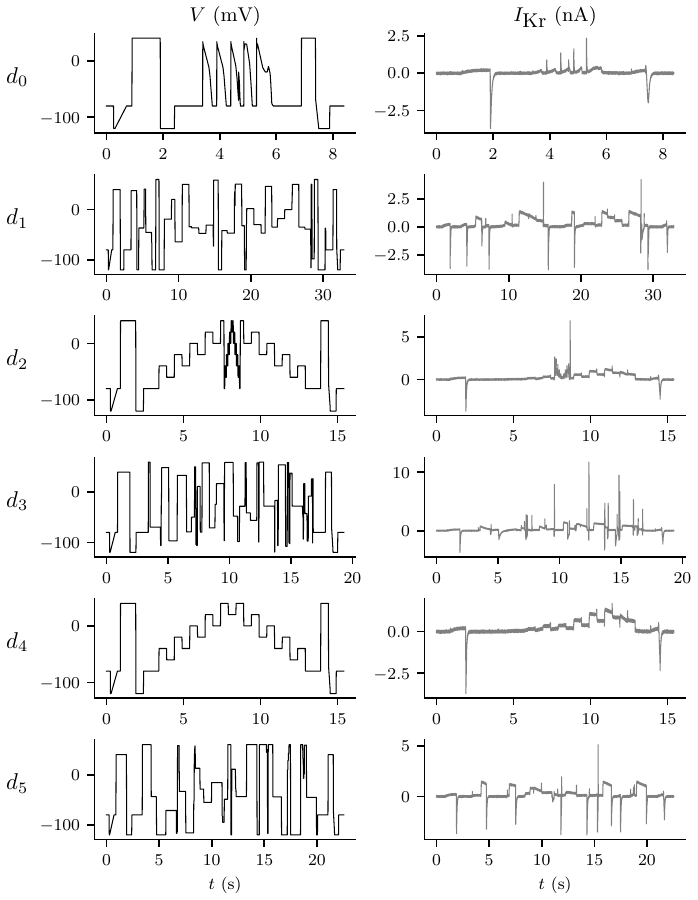}
    \caption{\textbf{Left}: a range of different input voltage-clamp protocols (forcing functions) used in this study. 
    \textbf{Right}: corresponding synthetic output data  I\textsubscript{Kr} simulated using the Beattie model with noise added as described in Section~\ref{sec:computational_methods}. 
    Here, we generate and plot data observed at a 10\,kHz sampling rate.
    Training protocols (all protocols except \(d_0\)) were tested for numerical identifiability \citep{fink_markov_2009}: inverse problems performed on synthetic data with repeatedly sampled random noise yielded parameter estimates with little variability.}
    \label{fig:protocols}
\end{figure}

A large amount of data can be recorded in voltage-clamp electrophysiology experiments: the current can be recorded at a several-kHz sampling rate, for many minutes.
In what follows, we take observations of the current at the same 10\,kHz frequency for all protocols.
Experimenters have a great deal of flexibility when it comes to specifying voltage-clamp protocol designs. 
We have published a number of studies on the benefits of `information-rich' experimental designs for these protocols, focusing on short protocols which explore a wide range of voltages and timescales \citep{beattie_sinusoidal_2018, lei_rapid_2019,lei_rapid_2019-1,clerx_four_2019,kemp_electrophysiological_2021}. 
In a real patch-clamp experiment, the amount of data we can obtain from each cell is limited. Hence, it is not feasible to perform many long protocols in sequence on the same cell. 
For this reason, we use six short information-rich protocols, denoted $d_0$ to $d_5$, as shown in Fig.~\ref{fig:protocols}. 

Here, we use simple designs consisting of a combination of sections where the voltage is held constant or `ramps' where the voltage increases linearly with time for compatibility with automated high-throughput patch clamp machines which are restricted to protocols of this type.
For the protocols included in this paper, short identical sequences including ramps are placed at the beginning and end of each protocol. 
In real experiments, these elements will allow for quality control, leak subtraction, and the estimation of the reversal potential \citep{lei_rapid_2019,lei_rapid_2019-1}. 
The central portion, consisting of steps during which the voltage is held constant, is what varies between protocols.

Not all possible designs are suitable for training models. 
Sometimes we encounter protocols for which distant pairs of parameter sets yield approximately equal model output---i.e.\ the model output for a protocol is not sensitive to certain (possibly large) changes in the model parameters.
Subsequently, when training the model to data generated from this protocol, many different parameter sets give similar fits that are almost equally plausible.
This problem is loosely termed \emph{numerical unidentifiability} \citep{fink_markov_2009} and  generally speaking is best avoided, unless the resulting uncertainty in the model parameters is known to be immaterial regarding any possible future context of use.

For both the Beattie and Wang models, numerical unidentifiability is a problem for design \(d_0\)
(data not shown, but this phenomenon is illustrated for a similar I\textsubscript{Kr} model and protocol in Fig.~3 of \citet{whittaker_calibration_2020}).
Yet \(d_0\) mimics the transmembrane voltage of a heart cell in a range of scenarios, and so provides a good way to validate whether our models recapitulate well-studied, physiologically-relevant behaviour. 
In particular, the central portion of this voltage-protocol consists of a sequence of wave-forms, each of which resembles the action potential of muscle cells found in the heart \citep{ten_tusscher_model_2004}.
So in this study we use \(d_0\) as a validation protocol, but do not use it as a protocol for training models.

The remaining designs, \(d_1\)--\(d_5\), were constructed using various criteria under constraints on voltage ranges and the duration of each step. The design
$d_1$ was designed algorithmically by sampling from a probability distribution placed over possible parameter sets and maximising the difference in model outputs between all pairs of parameter sets sampled from this distribution; $d_5$ was the result of the same algorithm applied to the Wang model.
In contrast, $d_4$ is a manual design we have used previously \citep{lei_rapid_2019} based on a simplification of a sinusoidal design \citep{beattie_sinusoidal_2018}.
The design, $d_2$ is a further manual refinement of $d_4$ to explore inactivation processes (rapid loss of current at high voltages) more thoroughly. Finally, 
$d_3$ is based on maximising the exploration of the model phase-space for the Beattie model, visiting as many combinations of binned model states and voltages as possible.
The main thing to note for this study however, is that \(d_1\)--\(d_5\) result in good numerical identifiability \citep{fink_markov_2009} for both models---that is, when used in synthetic data studies that attempt to re-infer the underlying parameters, all five protocol designs yield very low-variance parameter estimates (as shown in Fig.~\ref{tab:CaseI_parameters} for $\lambda=1$ for the Beattie model, and Fig.~\ref{tab:CaseII_Wang_parameters} for the Wang model).
This is a useful property, because it allows us to disregard the (very small) effect of different random noise in the synthetic data on the spread of our predictions (Equation~\ref{eqn:spread_of_prediction_interval}). 


\subsubsection{Computational Methods} \label{sec:computational_methods}
\paragraph{Numerical solution of ODEs}

Any time we calculate \(\mathbf{y}(\btheta; d)\), we must solve a system of ordinary differential equations.
We use a version of the LSODA solver designed to work with the Numba package, and Python to allow for the generation of efficient just-in-time compiled code from symbolic expressions. 
We partitioned each protocol into sections where the voltage is constant or changing linearly with respect to time because this sped up our computations.
We set LSODA's absolute and relative solver tolerances to \(10^{-8}\). 
The fact that the total number of channels is conserved in our models, allows us to reduce the number of ODEs we need to solve from \(N\) to \(N - 1\) \citep{fink_markov_2009}. 

\paragraph{Synthetic data generation}
Having computed the state of the system at each observation time, \((\mathbf{x}(t_i, \btheta^*, d))_{i=1}^{n_d}\), it is simple to compute \(y_i\) by substituting \(\mathbf{x}\) into our observation function (Equation~\ref{eqn:general_observation}). 
Finally, to add noise, we obtain \(n_d\) independent samples using Equation~\ref{eqn:general_observation}, 
using \emph{NumPy}'s \citep{harris2020array} interface to the PCG-64 pseudo-random number generator. 
Here, because we are using equally spaced observations with a 10\,kHz sampling rate, \(n_d = 10,000  \times t_\text{duration}\) where \(t_\text{duration}\) is the length of the protocol's voltage trace in seconds. 

\paragraph{Optimisation}

Finding the least-squares estimates, or, equivalently, minimising Equation~\ref{eqn:rmse} is (in general) a nonlinear optimisation problem for which there exist many numerical methods. 
We use CMA-ES \citep{hansen_cma_2016} as implemented by the PINTS interface \citep{clerx_probabilistic_2019}.
CMA-ES is a global, stochastic optimiser that has been applied successfully to many similar problems.

We follow the optimisation advice described in \citet{clerx_four_2019}. 
That is, for parameters `$a$' and `$b$' in state transition rates of the form $k=a \exp{(bV)}$, the optimiser works with `\(\log a\)' and untransformed `$b$' parameters. 
We enforce fairly lenient constraints on our parameter space, \(\Theta\), to prevent a proposed parameter set from forcing transitions to become so fast/slow that the ODE system becomes very stiff and computationally difficult to solve.
In particular, we take a similar approach to \citet{clerx_four_2019} we require that every parameter is positive, and, for ease of computation,
\begin{equation}
    1.67\times10^{-5} \text{\,ms}^{-1} \leq k_\text{max} \leq 10^{3}\text{\,ms}^{-1} \label{eqn:trapezium_prior_constaint}
\end{equation} where \(k_\text{max}\) is the maximum transition rate, \(k(V)\), for all \[V \in [-120\text{\,mV}, +60\text{\,mV}],\] which is the voltage range used in our protocols (Fig.~\ref{fig:protocols}).

Because CMA-ES is a stochastic algorithm, repeated runs can produce different output. To ensure that we have found the global minimum (Equation~\ref{eqn:parameter_estimation}), we repeat every optimisation numerous times (25 repeats for \(\lambda=1\) in Case I, \(5\) repeats for subsequent \(\lambda\), and \(25\) repeats in Case II). 
 
Moreover, in Section~\ref{sec:caseII}, when training the discrepant model, our initial guesses for the kinetic parameters were randomly sampled using
\begin{equation}
    \log_{10}(p) \sim U(-7, -1),
\end{equation} 
whereas we set the maximal conductance initial guess (which only affects the observation function) to the value used for data generation (even though these data were generated using a different model structure). 
We then check that our parameter set satisfies Equation~\ref{eqn:trapezium_prior_constaint}, and resample if necessary before commencing the optimisation routine.

The examples presented in Sections~\ref{sec:CaseI} and \ref{sec:caseII} require the solution of many optimisation problems. 
For speed, these tasks may be organised in such a way that multiple optimisation problems can be solved in parallel. 



\section{Results}\label{sec:results}

In this section, we use synthetically generated data to explore two cases of model discrepancy in Markov models of I\textsubscript{Kr}. 
In this first case, we gradually introduce discrepancy into a model with the correct structure by fixing one of its parameters to values away from the DGP parameter set. 
Then, in Section~\ref{sec:caseII}, we apply the same methods to another case where the model structure is incorrectly specified. 
In both cases, we take a literature model of I\textsubscript{Kr} together with Gaussian IID noise to be the DGP.

\subsection{Case I: Misspecified maximal conductance}
\label{sec:CaseI}

In this case, we assume a correctly specified model, but assume  increasingly erroneous values of one particular parameter and investigate how this impacts the protocol-dependence our parameter estimates and the predictive accuracy of our models. 
Also, we explore how the spread in our model predictions (Equation~\ref{eqn:spread_of_prediction_interval}) increases as the amount of discrepancy increases (in a particular manner).

To do this, we simulate data generation from each training protocol, as outlined, ten times using Gaussian IID noise with standard deviation (0.03nA). Specifically, we take the true DGP to be the Beattie model, as shown in Fig.~\ref{fig:beattie_wang_diagrams}.  Then, we fix the maximal conductance (\(g\)) to a range of values, and infer the remaining model parameters from the synthetic data, generated using the true parameter set, \(\btheta^*\). We assume that the standard deviation of the Gaussian noise is known because it can be well estimated from the initial portion of each protocol where the current is stationary.

When training our models, we use a restriction of the usual parameter space to fit the data by assuming some fixed value, \(\lambda\), for the maximal conductance, \(g\). 
In this way, we reformulate the optimisation problem slightly such that Equation~\ref{eqn:parameter_estimation} becomes
\begin{equation}
\hat\btheta_{\lambda}(d) = \text{argmin}_{\btheta\in\Theta_\lambda} \left\{\text{RMSE}\left(\mathbf y(\btheta ; d), \mathbf{z}(d)\right)\right\},
\end{equation}  
where \(\Theta_\lambda\) is the subset of parameter space where the maximal conductance is fixed to \(\lambda g\).
For each repeat of each protocol, we solve this optimisation problem for each  scaling factor, \(\lambda \in \left\{\frac{1}{4}, \frac{1}{2}, 1, 2, 4 \right\}\). 
These calculations are identical to those used in the computation of \emph{profile likelihoods} under the assumption of additive IID Gaussian errors \citep{bates_nonlinear_1988}.

Next, we fit these restricted parameter-space models to the same dataset and assess their predictive power under the application of a validation protocol. 
We do this for each possible pair of training and validation protocols.

To reduce the time required for computation we fit our discrepant models sequentially, starting at \(\lambda=1\) and increasing or decreasing \(\lambda\), using previous parameter estimates as an initial guess. 
This is done so that, for example, the kinetic parameters found by fixing \(\lambda = 2\) are used as our initial guess when we fit the model with \(\lambda = 4\), unless the original kinetic parameters (Table~\ref{tab:parameter_sets}) provide a lesser RMSE than the results of the previous optimisation. 

\begin{figure}[htbp]
    \centering
    \includegraphics{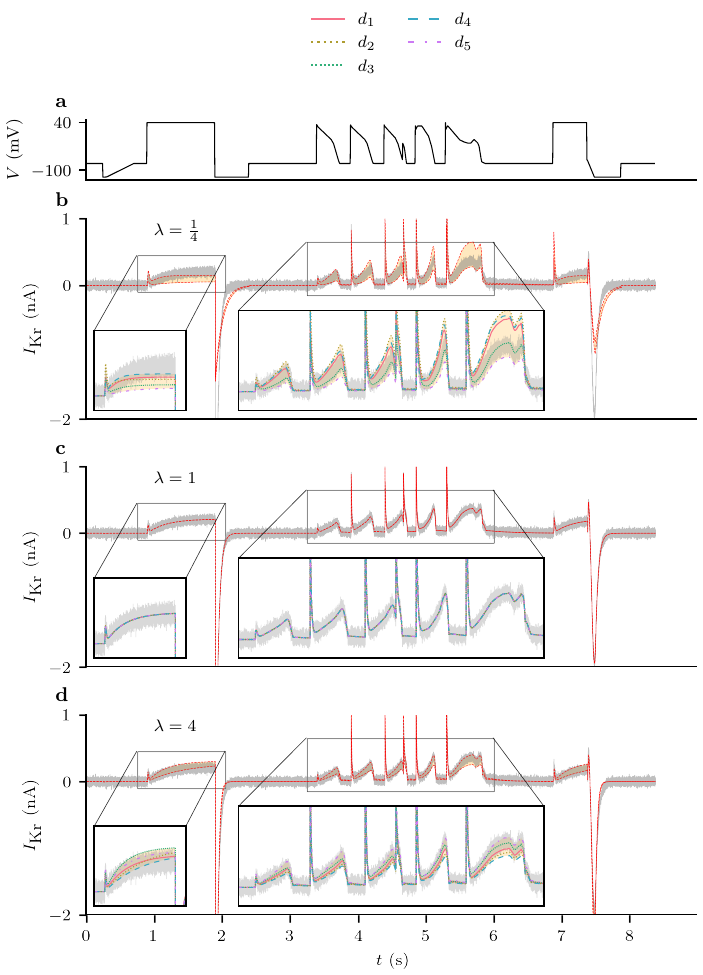}
    \caption{The set of predictions (Equation~\ref{eqn:set_of_predictions}) shown for parameter estimates obtained by training with different values of \(\lambda\) to synthetic data under \(d_1, \ldots, d_5\) (using the Beattie model). 
    The synthetically generated data used for model validation are shown in grey and and the spread of the predictions is highlighted in yellow.
    (\textbf{a}): the voltage trace for \(d_0\).
    (\textbf{b}): The set of predictions with \(\lambda = \frac 1 4\). 
    (\textbf{c}): the set of predictions with \(\lambda = 1\), that is, under the assumption of the correct  maximal conductance (\(g\)). 
    (\textbf{d}): the set of predictions with \(\lambda = 4\). 
    N.B. the `angular' nature of the current is not a plotting artefact, but reflects the fact the voltage clamp (\textbf{a}) is constructed from a series of linear ramps for compatibility with automated voltage clamp machines.
    }
    \label{fig:example_1_prediction_plots}
\end{figure}

\begin{figure}[htbp]
    \centering
    \includegraphics{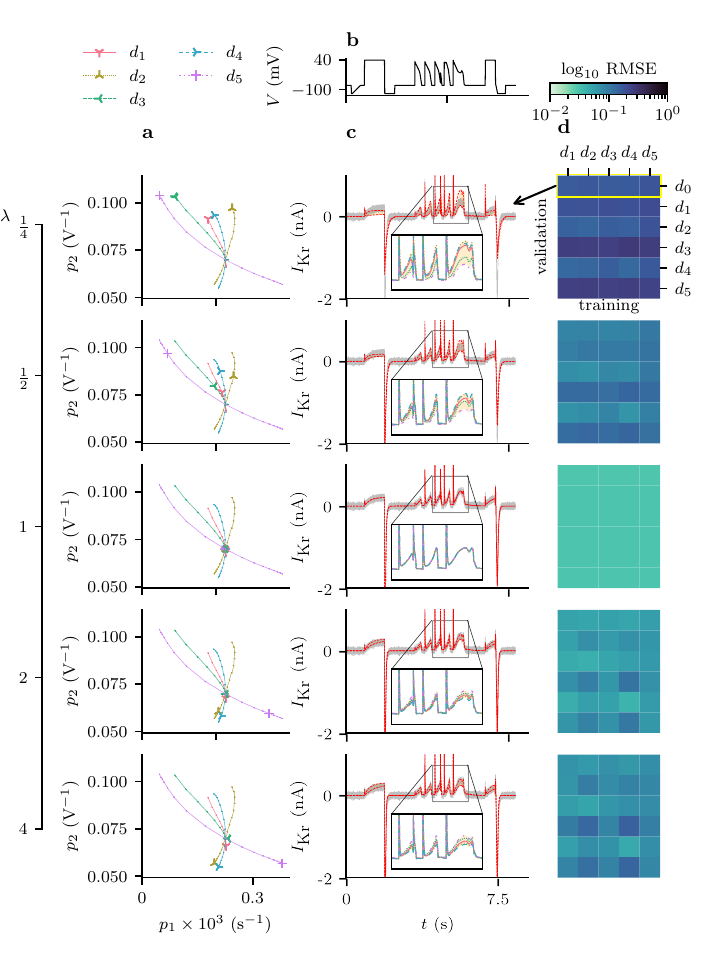}
    \caption{Discrepancy in parameter estimates and subsequent currents when a non-discrepant model is fitted to synthetic data, with all parameters free except the maximal conductance, \(g\), which is scaled by some factor \(\lambda\), ($g=\lambda g^*$), where $g^*$ is the true value. 
    (\textbf{a}):  estimates of \(\theta_1\) and \(\theta_2\) obtained by training with different protocols for 10 repeats of the DGP. 
    The lines (linearly interpolated using 17 values for \(\lambda 
    \in [
    \frac{1}{4}, 4] \)) show how the estimates from each protocol improve as \(\lambda\rightarrow 1\). 
    (\textbf{b}): $d_0$ voltage protocol. 
    (\textbf{c}): the spread of predictions of \(I_\text{Kr}\) under the $d_0$ protocol using the parameter estimates in Column \textbf{a}. 
    (\textbf{d}): a heatmap showing the predictive error obtained by training and validating for each pair of protocols. Here \textbf{c} corresponds with the top row of each heatmap, as indicated.
    }
    \label{fig:example_1_results}
\end{figure}

The spread in predictions for the validation protocol, \(d_0\), for, \(\lambda \in \{\frac 1 4, 1 , 4\)\} is shown in Fig.~\ref{fig:example_1_prediction_plots}. 
A more complete summary of these results is provided by Fig.~\ref{fig:example_1_results}. 
Here, when \(\lambda = 1\) (the central row of Fig.~\ref{fig:example_1_results}), we can see that no matter what protocol is used to train the model, the distribution of parameter estimates (panel \textbf{a}) is centred around their true values, and the resultant predictions are all accurate (Fig.~\ref{fig:example_1_results}, panels \textbf{b} and \textbf{c}). 
In contrast, when the maximal conductance, \(g\) is set to an incorrect value our parameter estimates become biased, and overall, our predictions become much less accurate. 
This effect on predictive accuracy is also shown in Fig.~\ref{fig:S1}. 

Moreover, the inaccuracy in our parameter estimates and our predictions varies depending on the design used to train the model. 
This effect does not appear to be symmetrical, with  \(\lambda < 1\) seemingly resulting in more model discrepancy than  \(\lambda > 1\).

Further results are provided in Section~\ref{sec:appendix_B}. Figure~\ref{fig:S1} shows the error in our predictions of \(d_0\) as \(\lambda\) varies, Table~\ref{tab:CaseI_parameters} examines the distribution of our parameter estimates for each protocol (under repeated samples of the DGP) for different values of \(\lambda\) and Table~\ref{tab:CaseI_summary} shows the behaviour of our spread-of-predictions interval (Equation~\ref{eqn:spread_of_prediction_interval}) and midpoint prediction Equation~\ref{eqn:midpoint_prediction} for different values of \(\lambda\).

\subsection{Case II: Misspecified dynamics} \label{sec:caseII}
Next, we apply these methods to an example where we have misspecified the dynamics of the model (the function \(\mathbf f\)). 
We use two competing Markov models of hERG kinetics,
the Beattie model \citep{beattie_sinusoidal_2018}, and the Wang model \citep{wang_quantitative_1997}. 
These models have differing structures and differing numbers of states, as shown in Fig.~\ref{fig:beattie_wang_diagrams}. 
We generate a synthetic dataset under the assumption of the Wang model with Gaussian IID noise (with standard deviation \(0.03\)nA) and the original parameter set as given in \citeauthor{wang_quantitative_1997}, for all the protocols shown in Fig.~\ref{fig:protocols}. As in Case I, we assume the standard deviation of this noise is known.

Then, we are able to fit both models to this dataset, obtaining an ensemble of parameter estimates and performing cross-validation as described in Section~\ref{sec:methods}. 
In this way, we can assess the impact of choosing the wrong governing equations (the choice of \(\mathbf f\) in Equation~\ref{eqn:general_ode}), and its impact on the predictive accuracy of the model. 
We do this to investigate whether the techniques introduced in Section~\ref{sec:general_method} are able to provide some useful quantification of model discrepancy when the dynamics of I\textsubscript{Kr} are misspecified.

\begin{figure}[htbp]
    \centering
    \includegraphics{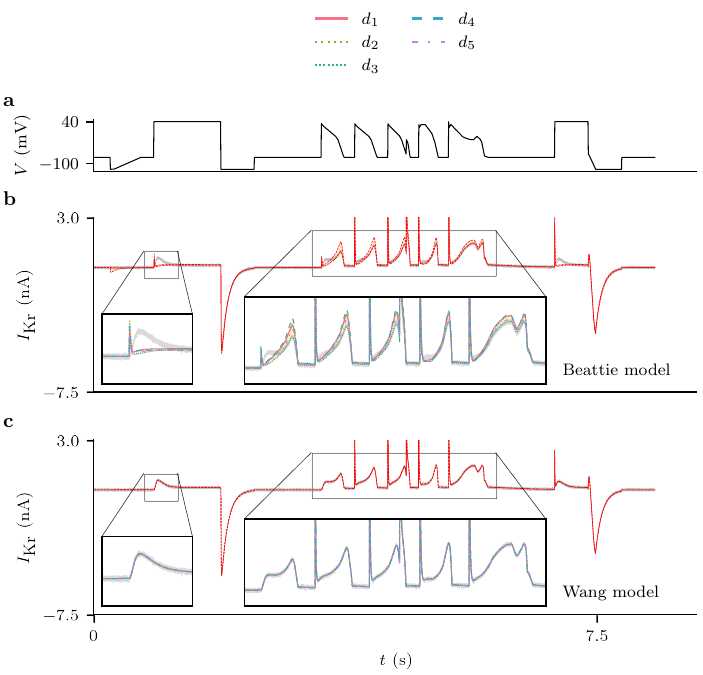}
    \caption{Case II: the set of predictions (Equation~\ref{eqn:set_of_predictions}) shown for parameter estimates obtained by training Beattie and Wang models with data synthetically generated using the Wang model. 
    (\textbf{a}): the \(d_0\) voltage-clamp protocol. (\textbf{b}): the set of predictions using the Beattie model. 
    (\textbf{c}): the set of predictions with Wang model, that is, with under the assumption of the correct model structure.}
    \label{fig:example_2_prediction_plots}
\end{figure}

\begin{figure}[htbp]
    \centering
 \includegraphics{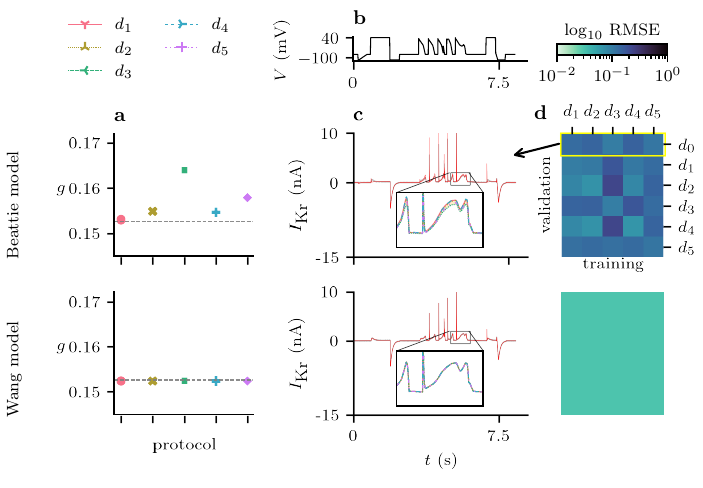}
    \caption{Model discrepancy between the Beattie model and synthetic data generated using the Wang model. 
    (\textbf{a}):  estimates of the maximal conductance obtained by training with different protocols for ten repeats of the DGP. 
    There is a noticeable protocol dependence for estimates obtained using the (discrepant) Beattie model, but the true underlying parameter (dashed line) can be accurately determined from any protocol when using the (correct) Wang model. 
    (\textbf{b}): $d_0$ voltage protocol. 
    (\textbf{c}): the spread of predictions for \(I_\text{Kr}\) under the $d_0$ protocol for discrepant (Beattie) and correct (Wang) models which are shown in more detail in Figure~\ref{fig:example_2_prediction_plots}.
    (\textbf{d}): cross-validation heatmaps for both the Beattie and Wang models fitted to this suite of protocols, averaged over ten repeated samples of the DGP for each protocol.
    }
    \label{fig:example2_main}
\end{figure}

Our results, presented in Figs.~ \ref{fig:example_2_prediction_plots} and \ref{fig:example2_main}, show how we expect a correctly specified model to behave in comparison to a discrepant model. 
We can see from the bottom row of Fig.~\ref{fig:example2_main}, that when training using the correctly specified derivative matrix, we were able to accurately recover the true maximal conductance using each and every protocol. 
Moreover, similarly to \textit{Case I}, no matter which protocol the correctly specified model was trained with, the resultant predictions were very accurate (as can be seen in Figure~\ref{fig:example_2_prediction_plots}).  

However, when the discrepant model was used, there was significant protocol dependence in our parameter estimates, and our predictions were much less accurate overall, but perhaps accurate for many applications. 
Moreover, it seems that for the majority of \(d_0\), the spread in predictions across training protocols (Equation~\ref{eqn:spread_of_prediction_interval}) was smaller than those seen in Case I, but there are certain portions where the discrepant model and DGP are noticeably different (as highlighted in Fig.~\ref{fig:example_2_prediction_plots}).
This may be due to the structural differences between the Wang and Beattie model. 
In particular, in the Wang model, channels transitioning from the high-voltage inactive state (\textit{I}), must transition through the conducting, open state (\textit{O}) in order to reach low-voltage closed states (\textit{C1}, \textit{C2}, \textit{C3}), causing a spike of current to be observed. 
Instead, channels in the Beattie model may transition through the inactive-and-closed state (\textit{IC}) on their way between \textit{O} and \textit{C}, resulting in reduced current during steps from high voltage to low voltage. 

\begin{figure}[htbp]
    \centering
    \includegraphics{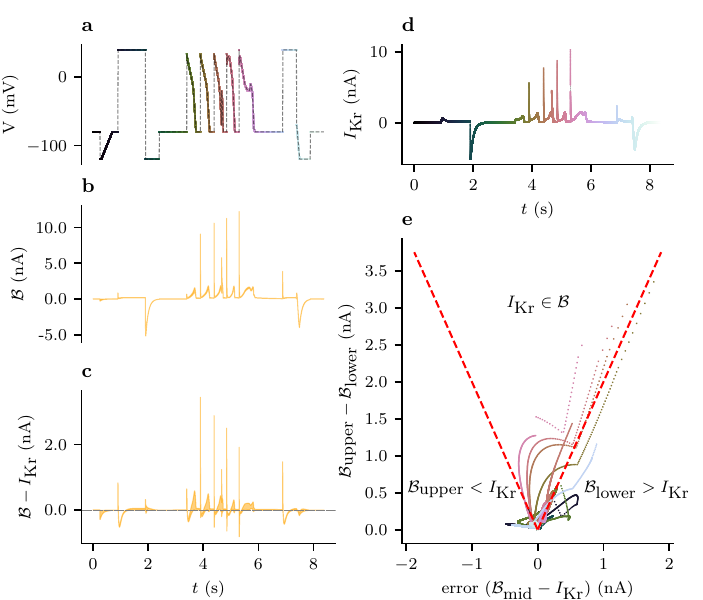}
    \caption{The spread in predictions obtained from different protocols provides a useful indicator of model discrepancy for \textit{Case II}. 
    (\textbf{a}): the validation voltage-clamp protocol, \(d_0\), with colours corresponding to panels \textbf{d} and \textbf{e}.
    (\textbf{b}): the spread-of-predictions interval (Equation~\ref{eqn:spread_of_prediction_interval}) for \(d_0\) using the Beattie model trained with \(d_1, \ldots, d_5\).
    (\textbf{c}): the true DGP subtracted from the spread-in-predictions interval.
    (\textbf{d}): the true DGP with the colour of each observation corresponding to panels \textbf{a} and \textbf{e}.
    (\textbf{e}): a scatter plot of the midpoint prediction (Equation~\ref{eqn:midpoint_prediction}) and the width of the predictive interval (Equation~\ref{eqn:spread_of_prediction_interval}) for every observation in \(d_0\). 
    Here, the red, dashed lines show the true value of \(I_\text{Kr}\) lies on the extremes of the range of predictions. 
    Accordingly, points above these lines show the observations for which the DGP lies inside this range, and the points below the line correspond to observations for which the true DGP lies outside this range. 
    The colours of these points correspond to those in panels \textbf{a} and \textbf{d}.}
    \label{fig:example2_error_compare}
\end{figure}

Nevertheless, our methods provide a useful indication of this model discrepancy. Fig.~\ref{fig:example2_error_compare}, examines the behaviour of our prediction interval (Equation~\ref{eqn:spread_of_prediction_interval}) in more detail. 
Importantly, we can see that our interval shows little uncertainty during sections of the protocol where there is little current (this is also seen in Fig.~\ref{fig:example2_main}\textbf{a} and Fig.~\ref{fig:example_2_prediction_plots}). This is ideal, because no reasonable model would predict a sizeable current here. On the other hand, we see that our intervals show significant uncertainty around the spikes in current that occur at the start of each action-potential waveform.
This is to be expected because it is known that these sections of the current time-series are particularly sensitive to differences in the `rapid inactivation' process in these models \citep{clerx_four_2019}. 

Further results regarding Case II are provided in Section~\ref{sec:appendix_C}. Table~\ref{tab:CaseII_Beattie_parameters} and Table~\ref{tab:CaseII_Wang_parameters} summarise the behaviour of our parameter estimates for each choice of model.

\FloatBarrier
\section{Discussion}\label{sec:discussion}


We have introduced an uncertainty quantification (UQ) approach to highlight when discrepancy is affecting model predictions. 
We demonstrated the use of this technique by providing insight into the effects of model discrepancy on a model of I\textsubscript{Kr} in electrically excitable cells. 
Here, we saw that under synthetically constructed examples of model discrepancy, there was great variability between the parameter estimates obtained using different experimental designs.
This variability is a consequence of the different compromises that a discrepant model has to make to fit different regimes of a true DGP's behaviour.
Consequently, these parameter estimates produced a wide range of behaviour during validation, despite each individual parameter estimate having little variability under repeated samples of the DGP. 

The variability in the model predictions stemming from this ensemble of parameter estimates is, therefore, an empirical way of characterising the predictive uncertainty due to model discrepancy. 
Usefully, our spread-of-prediction intervals (Equation~\ref{eqn:spread_of_prediction_interval}) correctly indicated little uncertainty when the ion channel model was exhibiting simple dynamics decaying towards a steady state, but more uncertainty during more complex dynamics, which was indeed when the largest discrepancies occurred. 
For many observations under our validation protocol, the true, underlying DGP lay inside this interval, indicating that Equation~\ref{eqn:spread_of_prediction_interval} may provide a useful indication of predictive error under unseen protocols. 
We expect that the presented methods may be of use for problems where the variability in parameter estimates (from repetitions of each individual protocol) is smaller than the variability between parameter estimates obtained from different protocols---because there is little noise, and lots of observations for example. 
In such cases, the variability in the extremes of our ensembles (\(\mathcal{B}_\text{upper}\) and \(\mathcal{B}_\text{lower}\)) is immaterial compared to the width of the interval (\(\mathcal{B}_\text{upper} - \mathcal{B}_\text{lower}\)).




\begin{figure}[tbh]
    \centering
    \includegraphics{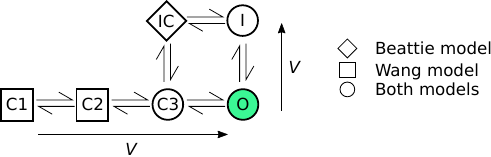}
    \caption{
    The Beattie and Wang models may be seen as special cases of this more complicated model. 
    The state labelled `C3' is called `C' in the Beattie model and `C3' in the Wang model. 
    The arrows outside the Markov state diagram indicate the direction in which rates increase with more positive voltage.
    }
    \label{fig:wang_beattie_combined}
\end{figure}



At first, Case I may seem like an artificial example---in practice, the maximal conductance is taken to be a model parameter and fitted along with the rest of the model. But Case I and Case II are similar: any two Markov models may be regarded as two special cases of a more general model with some transition rates pinned to \(0\) (as shown in Fig.~\ref{fig:wang_beattie_combined} for the models used in this paper, with some transition rate). Like in Case I, this means that different  model structures can be seen as restrictions of this larger model's parameter space.  Misspecified model structures can then be identified with subsets of parameter space which do not contain the true, data-generating parameter set (provided this larger model is structurally identifiable).

This means there is a setting in which Case II (misspecified governing equations) is an example of the type of discrepancy explored in Case I, where a ``true'' parameter value exists in the more general model, but is excluded in the parameter space being optimised over when training the model. 
This may prove a valuable perspective for modelling ion channel kinetics, where there are many candidate models \citep{mangold2021identification}, and each model may be seen as corresponding to some subset of a general model with a shared higher-dimensional parameter space. Model selection problems have been framed in this way previously \citep{akaike_information_1998,chen_network_2017}.



\subsection{Limitations}

Whilst the spread of predictions under some unseen protocol may provide some feasible range of predictions, we can see from Fig.~\ref{fig:example2_error_compare}, that our observables (the DGP without noise) often lie outside this range. 
This shown in Fig.~\ref{fig:example_2_prediction_plots}. 
Here, certain structural differences between the model and DGP may mean that the truth lies outside.
For this reason,  Equation~\ref{eqn:spread_of_prediction_interval} is best interpreted as a heuristic indication of predictive uncertainty stemming from model discrepancy, rather than providing any guarantees about the output of the DGP. 

Using more training protocols in the training set may increase the coverage of the DGP by our interval. 
Whilst the number of protocols that can be performed on a single biological cell is limited by time constraints \citep{beattie_sinusoidal_2018,lei_rapid_2019-1}, the utilisation of more protocols is likely preferable. 

Besides the types of discrepancies considered in Section~\ref{sec:results}, there are other ways that the DGP can differ from the fitted models.
For example, the DGP may not be accurately described by an ODE system, especially when ion channel numbers are small and the stochasticity of individual channels opening and closing is apparent.
In this circumstance, the  models can be cast in terms of stochastic differential equations (SDEs), as in  \citet{PhysRevE.83.041908}, and we can again consider an ensemble of parameter estimates (Equation~\ref{eqn:set_of_estimates}) and an ensemble of model predictions (Equation~\ref{eqn:set_of_predictions}.
The assumption of IID Gaussian errors for the observation noise model could also be inaccurate: auto-correlated noise processes \citep[e.g. as explored in][]{creswell2020noise,lambert2022autocorrelated}, or even experimental artefacts may be present, but all of these could be included in the modelled DGP \citep{lei_accounting_2020} and it remains to be seen how well our method would perform in these cases.

\subsection{Future Directions}




We were able to quantify model discrepancy by considering the predictive error of our models across a range of training and validation protocols. 
This provides a way of quantifying model discrepancy that can be compared across models, and could be used to select the most suitable model from a group of candidate models. 
For a given context of use, we suggest that the spread of predictions can be used to gauge the trustworthiness of a model's predictions. 
Even in a model which produces a plausible fit to each individual training protocol, a wide spread of predictions may indicate that a model is ill-suited to a particular predictive task, and should prompt careful reconsideration of the model and the experiments used for its training.
In this way, the disagreement between model predictions of \(d_0\) in Section~\ref{sec:CaseI} shows that the \(\lambda = \frac{1}{4}\) may not be suitable, owing to the relative width of this band of predictions, even in the absence of validation data, or knowledge of a more suitable model.

Our approach may provide insight into improved experimental design criteria \citep{lei_oed_2022}. 
Optimal experimental design approaches that assume knowledge of a correctly specified model may not be the most suitable in the presence of known discrepancy.
By adjusting these approaches to account for the uncertainty in choice of  model, we may be able to use these ideas to design experiments which allow for more robust model validation. 
One method would be to fit to data from a collection of training protocols, and to find a new protocol, for which the spread in ensemble predictions (Equation~\ref{eqn:spread_of_prediction_interval}) is maximised.

In this paper, we have applied our methodology to mathematical models of electrophysiology. 
In both the cases we considered, we saw that model discrepancy could lead to inaccurate predictive models, and due to the use of information-rich training protocols, the variability of our parameter estimates, under repeated samples of the DGP, was negligible. 
We propose that our methodology could be applied to similar problems, where there is high-frequency time-series data is available as well as mathematical model which are relatively accurate. 
There are many such biochemical reaction networks for which model selection remains a challenge, and there are numerous approaches to finding suitable mathematical models \citet{klimovskaia_sparse_2016}. 
We propose the methodology outlined in this paper may be used to quantify the discrepancy in such models. 

In these examples, we saw little variability due to noise in our parameter inference, and therefore in our ensemble prediction's spread-of-prediction intervals and/or midpoint predictions, as shown in Section~\ref{sec:appendix_B} and Section~\ref{sec:appendix_C}. 
However, in other cases where there are fewer observations and more observational noise (for example), it may be more suitable to consider an analogous distribution-based approach where we consider Bayesian posteriors of our parameters instead of point estimates (such as the maximum likelihood estimator we used in this paper).

\subsection{Concluding remarks}

The spread of predictions of our ensembles, based on training to data from multiple experimental designs, provides a good indication of possible predictive error due to model discrepancy.
Ultimately, whilst our ensemble approach is no substitute for a correctly specified model, it is a useful tool for quantifying model discrepancy, predicting the size and direction of its effects, and may guide further experimental design and model selection approaches.

\section*{Data Availability}
Open source code for all the simulation studies and plots in this paper can be found at \url{https://github.com/CardiacModelling/empirical_quantification_of_model_discrepancy}.
A permanently archived version is available on Zenodo \url{https://doi.org/10.5281/zenodo.8409925}.

\section*{Acknowledgements}
This work was supported by the Wellcome Trust (grant no.~212203/Z/18/Z);
the Science and Technology Development Fund, Macao SAR (FDCT) [reference no.~0048/2022/A]; the EPSRC [grant no.~EP/R014604/1]; and the Australian Research Council [grant no.~DP190101758].
GRM acknowledges support from the Wellcome Trust via a Wellcome Trust Senior Research Fellowship to GRM.
CLL acknowledges support from the FDCT and support from the University of Macau via a UM Macao Fellowship.
We acknowledge Victor Chang Cardiac Research Institute Innovation Centre, funded by the NSW Government.
The authors would like to thank the Isaac Newton Institute for Mathematical Sciences for support and hospitality during the programme The Fickle Heart when some work on this paper was undertaken. 

This research was funded in whole, or in part, by the Wellcome Trust [212203/Z/18/Z]. 
For the purpose of open access, the authors have applied a CC-BY public copyright licence to any Author Accepted Manuscript version arising from this submission.

\newpage\clearpage
\bibliography{references}

\newpage\clearpage
\section{Appendices}
\subsection{Appendix A: Parameterisation of Markov models} \label{sec:appendix_A}
\subsubsection{Beattie model}
In full, the system of ODEs is, 
\begin{equation}
\dfrac{\textrm{d}\mathbf x}{\text{d}t} =  \left[
\begin{matrix}-k_1 -k_3& 0 & k_4 & k_2\\
0 & -k_2 - k_4 & k_1 & k_3\\
k_3 &k_2 &-k_1 - k_4 & 0\\
k_1 & k_4 & 0 & -k_2 - k_3\end{matrix}
\right] \mathbf x, \label{eqn:beattie_transition_matrix}
\end{equation}
 where 
\begin{align}
 k_{1} &= p_1e^{p_2 V},\\
 k_{2} &= p_3 e^{- p_4 V},\\
 k_{3} &= p_5 e^{p_6 V} \hspace{1em},\\ 
 k_{4} &= p_7 e^{- p_8 V }.
\end{align} 
Hence, the corresponding parameter set is,
\begin{equation}
 \btheta = \begin{bmatrix}
     p_1, \;
     \ldots, \;
     p_8, \;
     g
 \end{bmatrix}^T,
\end{equation}
and, 
\begin{align} 
\mathbf x = \begin{bmatrix}C,\;I,\;IC,\;O\end{bmatrix}^T. 
\end{align}

\subsubsection{Wang model}
 We may write this model's governing system of ODEs as
\begin{equation}
\dfrac{\textrm{d}\mathbf x}{\text{d}t} =  \left[
\begin{matrix}
- \alpha_{a0} & \beta_{a0} & 0 & 0 & 0
\\\alpha_{a0} & - \beta_{a0} - k_f & k_b & 0 & 0
\\0 & k_f & - k_b - \alpha_{a1} & \beta_{a1} & 0
\\0 & 0 & \alpha_{a1} & - \beta_{a1} - \alpha_1 & \beta_1
\\0 & 0 & 0 & \alpha_1 & -\beta_1
\end{matrix} \right] \mathbf x,
\label{eqn:wang_transition_matrix}
\end{equation}
 where 
\begin{align}
  a_{1} &= q_1 e^{q_2 V},\\ 
 a_{a0} &= q_3e^{q_4 V},\\
 a_{a1} &= q_5 e^{q_6 V},\\ 
 b_{a1} &= q_7 e^{- q_8 V},\\ 
 b_{1} &= q_9 e^{- q_{10} V} ,\\ 
 b_{a0} &= q_{11} e^{- q_{12} V }.\\ 
\end{align}  
The corresponding parameter set is, 
\begin{equation}\btheta = 
 \begin{bmatrix}
 q_1,\;\ldots,\; q_{12},\;k_f,\;k_b
 \end{bmatrix}^T,
\end{equation}
and 
\begin{equation} 
 \mathbf x = 
 \begin{bmatrix}
C_{1}, & C_{2},& C_{3}, & O, & I
 \end{bmatrix}^T.
\end{equation}

The default parameter values for both models are presented in Table \ref{tab:parameter_sets}. 
\begin{table}[bthp]
\centering
\begin{minipage}[t]{.49\linewidth}
\centering
\vspace{0pt} 
\begin{tabular}{c l l}
\multicolumn{3}{c}{Wang Model}\\
\toprule
Parameter & Value & Units \\ 
\midrule
 \(g\)   & \(1.52\times 10^{-1}\)&  \(\mu\)S \\
 \(k_b\)   & \(3.68\times 10^{-2}\)&ms\(^{-1}\)\\  
\(k_f\)   & \(2.38\times 10^{-2}\)&ms\(^{-1}\)\\  
 \(q_{1}\)   & \(9.08\times 10^{-2}\)&ms\(^{-1}\)\\
 \(q_{2}\)   & \(2.34\times 10^{-2}\)&mV\(^{-1}\)\\
\(q_3\)    & \(2.23\times 10^{-2}\)& ms\(^{-1}\)\\
 \(q_4\)   & \(1.18\times 10^{-2}\) & mV\(^{-1}\)\\
 \(q_5\)   & \(1.37\times10^{-2}\)&ms\(^{-1}\)\\
 \(q_6\)   & \(3.82\times 10^{-2}\)& mV\(^{-1}\)\\
 \(q_7\)   & \(6.89\times 10^{-5}\)& ms\(^{-1}\)\\
 \(q_8\)   & \(4.18\times 10^{-2}\)&mV\(^{-1}\)\\
 \(q_9\)   & \(6.50\times 10^{-3}\)&ms\(^{-1}\)\\
\(q_{10}\)   & \(3.27\times 10^{-2}\)&mV\(^{-1}\)\\
\(q_{11}\)   & \(4.70\times 10^{-2}\)&mV\(^{-1}\)\\
 \(q_{12}\)   & \(6.31\times 10^{-2}\)&ms\(^{-1}\)\\
\bottomrule
\end{tabular}
\end{minipage}
\begin{minipage}[t]{.49\linewidth}
\centering
\vspace{0pt}
\begin{tabular}{c l l}
\multicolumn{3}{c}{Beattie Model}\\
\toprule
Parameter & Value & Units \\ 
\midrule
 \(g\)   & \(1.52 \times 10^{-1}\)&  \(\mu\)S \\
\(p_1\)    & \(2.26\times 10^{-4}\)& ms\(^{-1}\)\\
 \(p_2\)   & \(6.99\times 10^{-2}\) & mV\(^{-1}\)\\
 \(p_3\)   & \(3.45\times 10^{-5}\)&ms\(^{-1}\)\\
 \(p_4\)   & \(5.46\times 10^{-2}\)&mV\(^{-1}\)\\
 \(p_5\)   & \(8.73\times 10^{-2}\)&ms\(^{-1}\)\\
 \(p_6\)   & \(8.91\times 10^{-3}\)& mV\(^{-1}\)\\
 \(p_7\)   & \(5.15\times 10^{-3}\)& ms\(^{-1}\)\\
 \(p_8\)   & \(3.16\times 10^{-2}\)&mV\(^{-1}\)\\
 \bottomrule
\end{tabular} 
\end{minipage} \vspace{.5em}
\caption{The default parameter sets we use for the \citet{wang_quantitative_1997} and \citet{beattie_sinusoidal_2018} models. 
The same maximal conductance (\(g\)) is used for both models.}
\label{tab:parameter_sets}
\end{table}

\subsection{Appendix B: further Case I results}  \label{sec:appendix_B}
\begin{figure}
     \centering
     \includegraphics{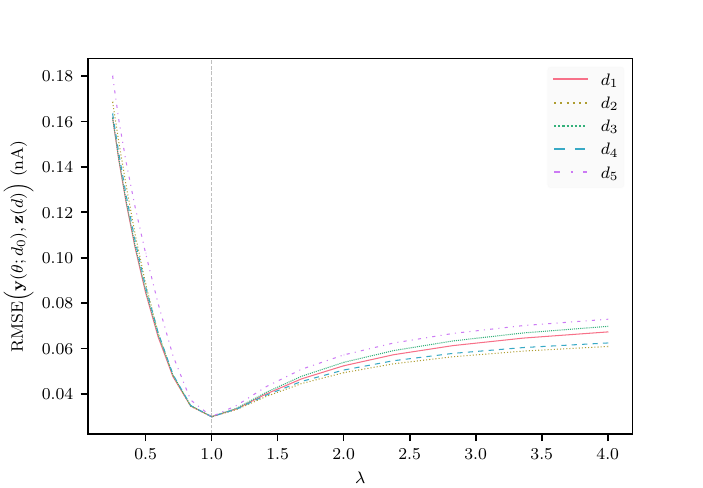}
     \caption{Case I: predictive accuracy (under our validation protocol, \(d_0\)) decreases as \(\lambda\rightarrow 1\). For \(17\) values of \(\lambda\) (\(\frac{1}{4} \leqslant  \lambda \leqslant 4\)), the predictive error (averaged over repeats) is shown for each training protocol (\(d_1\)--\(d_5\))
     }
     \label{fig:S1}
 \end{figure}

The predictive accuracy (under the validation protocol, \(d_0\)))  of the model used in Section~\ref{sec:caseII}, trained using each protocol, for a range of values of \(\lambda\) is shown in Fig.~\ref{fig:S1}. Here, we see that as there is more model discrepancy (when \(\lambda\) moves away from \(1\)) our predictions become less accurate.

The parameter estimates obtained in Sections~\ref{sec:CaseI} and \ref{sec:caseII} are summarised in Table~\ref{tab:CaseI_parameters}, Table~\ref{tab:CaseII_Beattie_parameters} and Table~\ref{tab:CaseII_Wang_parameters}, respectively. Here, we can see that when the model is misspecified, the small standard deviation in our estimates (across fits to different samples of our DGP) is small compared to the differences between estimates obtained from different protocols---the choice of training protocol is less important when there is no model discrepancy.

 Table~\ref{tab:CaseI_parameters} details the distribution of each parameter estimate (under repeated samples of the DGP) for each protocol as \(\lambda\) varies (as described in Section~\ref{sec:CaseI}. Whereas, Table~\ref{tab:CaseI_summary} shows how our spread-of-prediction intervals change under different values of \(\lambda\). Here, we can see that each parameter estimate, as well as \(\mathcal{B}\) itself, show little variability under repeated samples of the DGP.
 
 \begin{table}[]
     \centering
           \setlength{\tabcolsep}{5pt}
     \begin{tabular}{llccccc}
\toprule
$\lambda$ &  & $d_1$ & $d_2$ & $d_3$ & $d_4$ & $d_5$ \\
\midrule
\multirow[c]{8}{*}{0.25} & $p_1$ & 1.8E-04\(\pm\)2E-07 & 2.4E-04\(\pm\)7E-07 & 9.0E-05\(\pm\)1E-06 & 1.9E-04\(\pm\)4E-07 & 4.7E-05\(\pm\)8E-08 \\
 & $p_2$ & 9.1E-02\(\pm\)5E-05 & 9.7E-02\(\pm\)9E-05 & 1.0E-01\(\pm\)2E-04 & 9.3E-02\(\pm\)8E-05 & 1.0E-01\(\pm\)4E-05 \\
 & $p_3$ & 3.0E-05\(\pm\)2E-08 & 1.9E-05\(\pm\)3E-08 & 1.2E-05\(\pm\)4E-08 & 1.7E-05\(\pm\)4E-08 & 9.3E-06\(\pm\)2E-08 \\
 & $p_4$ & 4.8E-02\(\pm\)6E-06 & 5.2E-02\(\pm\)2E-05 & 5.5E-02\(\pm\)3E-05 & 5.3E-02\(\pm\)2E-05 & 5.8E-02\(\pm\)2E-05 \\
 & $p_5$ & 5.0E-02\(\pm\)5E-05 & 5.5E-02\(\pm\)6E-05 & 5.0E-02\(\pm\)5E-05 & 7.9E-02\(\pm\)3E-04 & 5.1E-02\(\pm\)8E-05 \\
 & $p_6$ & 1.5E-02\(\pm\)2E-05 & 1.0E-02\(\pm\)4E-05 & 1.0E-02\(\pm\)2E-05 & 1.8E-02\(\pm\)5E-05 & 1.3E-02\(\pm\)3E-05 \\
 & $p_7$ & 1.5E-02\(\pm\)2E-05 & 1.3E-02\(\pm\)2E-05 & 9.8E-03\(\pm\)2E-05 & 2.3E-02\(\pm\)7E-05 & 1.0E-02\(\pm\)5E-05 \\
 & $p_8$ & 4.4E-02\(\pm\)2E-05 & 4.6E-02\(\pm\)5E-05 & 5.1E-02\(\pm\)4E-05 & 3.6E-02\(\pm\)4E-05 & 5.8E-02\(\pm\)1E-04 \\ \midrule
\multirow[c]{8}{*}{0.50} & $p_1$ & 2.2E-04\(\pm\)2E-07 & 2.5E-04\(\pm\)6E-07 & 2.0E-04\(\pm\)6E-07 & 2.1E-04\(\pm\)4E-07 & 6.9E-05\(\pm\)2E-07 \\
 & $p_2$ & 7.6E-02\(\pm\)3E-05 & 8.5E-02\(\pm\)8E-05 & 8.0E-02\(\pm\)6E-05 & 8.8E-02\(\pm\)7E-05 & 9.7E-02\(\pm\)6E-05 \\
 & $p_3$ & 3.7E-05\(\pm\)2E-08 & 3.4E-05\(\pm\)5E-08 & 3.6E-05\(\pm\)6E-08 & 3.6E-05\(\pm\)6E-08 & 2.5E-05\(\pm\)3E-08 \\
 & $p_4$ & 5.0E-02\(\pm\)4E-06 & 5.1E-02\(\pm\)1E-05 & 5.0E-02\(\pm\)2E-05 & 5.1E-02\(\pm\)1E-05 & 5.3E-02\(\pm\)9E-06 \\
 & $p_5$ & 7.5E-02\(\pm\)7E-05 & 7.5E-02\(\pm\)6E-05 & 6.8E-02\(\pm\)6E-05 & 8.4E-02\(\pm\)2E-04 & 7.7E-02\(\pm\)7E-05 \\
 & $p_6$ & 9.4E-03\(\pm\)1E-05 & 9.0E-03\(\pm\)3E-05 & 9.7E-03\(\pm\)2E-05 & 1.1E-02\(\pm\)3E-05 & 9.9E-03\(\pm\)1E-05 \\
 & $p_7$ & 9.3E-03\(\pm\)1E-05 & 8.2E-03\(\pm\)7E-06 & 6.8E-03\(\pm\)8E-06 & 1.0E-02\(\pm\)3E-05 & 8.5E-03\(\pm\)1E-05 \\
 & $p_8$ & 3.7E-02\(\pm\)1E-05 & 3.8E-02\(\pm\)2E-05 & 4.0E-02\(\pm\)1E-05 & 3.5E-02\(\pm\)3E-05 & 3.8E-02\(\pm\)2E-05 \\ \midrule
\multirow[c]{8}{*}{1.00} & $p_1$ & 2.3E-04\(\pm\)2E-07 & 2.3E-04\(\pm\)4E-07 & 2.3E-04\(\pm\)6E-07 & 2.3E-04\(\pm\)3E-07 & 2.3E-04\(\pm\)6E-07 \\
 & $p_2$ & 7.0E-02\(\pm\)3E-05 & 7.0E-02\(\pm\)7E-05 & 7.0E-02\(\pm\)5E-05 & 7.0E-02\(\pm\)5E-05 & 7.0E-02\(\pm\)6E-05 \\
 & $p_3$ & 3.4E-05\(\pm\)1E-08 & 3.4E-05\(\pm\)4E-08 & 3.4E-05\(\pm\)5E-08 & 3.4E-05\(\pm\)5E-08 & 3.4E-05\(\pm\)2E-08 \\
 & $p_4$ & 5.5E-02\(\pm\)5E-06 & 5.5E-02\(\pm\)1E-05 & 5.5E-02\(\pm\)1E-05 & 5.5E-02\(\pm\)1E-05 & 5.5E-02\(\pm\)5E-06 \\
 & $p_5$ & 8.7E-02\(\pm\)7E-05 & 8.7E-02\(\pm\)8E-05 & 8.7E-02\(\pm\)6E-05 & 8.7E-02\(\pm\)3E-04 & 8.7E-02\(\pm\)6E-05 \\
 & $p_6$ & 8.9E-03\(\pm\)9E-06 & 8.9E-03\(\pm\)2E-05 & 8.9E-03\(\pm\)1E-05 & 8.9E-03\(\pm\)3E-05 & 8.9E-03\(\pm\)1E-05 \\
 & $p_7$ & 5.2E-03\(\pm\)6E-06 & 5.2E-03\(\pm\)4E-06 & 5.2E-03\(\pm\)4E-06 & 5.1E-03\(\pm\)1E-05 & 5.2E-03\(\pm\)4E-06 \\
 & $p_8$ & 3.2E-02\(\pm\)9E-06 & 3.2E-02\(\pm\)2E-05 & 3.2E-02\(\pm\)8E-06 & 3.2E-02\(\pm\)3E-05 & 3.2E-02\(\pm\)1E-05 \\
\multirow[c]{8}{*}{2.00} & $p_1$ & 2.3E-04\(\pm\)2E-07 & 2.1E-04\(\pm\)3E-07 & 2.3E-04\(\pm\)7E-07 & 2.1E-04\(\pm\)3E-07 & 3.4E-04\(\pm\)6E-07 \\
 & $p_2$ & 6.8E-02\(\pm\)3E-05 & 6.1E-02\(\pm\)7E-05 & 7.0E-02\(\pm\)6E-05 & 5.8E-02\(\pm\)5E-05 & 6.0E-02\(\pm\)4E-05 \\
 & $p_3$ & 3.0E-05\(\pm\)1E-08 & 3.0E-05\(\pm\)4E-08 & 2.5E-05\(\pm\)4E-08 & 2.8E-05\(\pm\)5E-08 & 3.4E-05\(\pm\)2E-08 \\
 & $p_4$ & 5.9E-02\(\pm\)5E-06 & 5.8E-02\(\pm\)1E-05 & 6.0E-02\(\pm\)1E-05 & 5.8E-02\(\pm\)2E-05 & 5.8E-02\(\pm\)6E-06 \\
 & $p_5$ & 9.0E-02\(\pm\)7E-05 & 8.9E-02\(\pm\)8E-05 & 9.5E-02\(\pm\)7E-05 & 9.5E-02\(\pm\)3E-04 & 9.0E-02\(\pm\)5E-05 \\
 & $p_6$ & 1.0E-02\(\pm\)6E-06 & 9.6E-03\(\pm\)2E-05 & 9.0E-03\(\pm\)1E-05 & 8.6E-03\(\pm\)4E-05 & 9.7E-03\(\pm\)8E-06 \\
 & $p_7$ & 2.6E-03\(\pm\)3E-06 & 2.9E-03\(\pm\)2E-06 & 2.9E-03\(\pm\)2E-06 & 2.8E-03\(\pm\)1E-05 & 2.7E-03\(\pm\)2E-06 \\
 & $p_8$ & 2.7E-02\(\pm\)8E-06 & 2.7E-02\(\pm\)2E-05 & 2.6E-02\(\pm\)6E-06 & 2.8E-02\(\pm\)3E-05 & 2.7E-02\(\pm\)8E-06 \\ \midrule
\multirow[c]{8}{*}{4.00} & $p_1$ & 2.3E-04\(\pm\)2E-07 & 2.0E-04\(\pm\)3E-07 & 2.3E-04\(\pm\)8E-07 & 2.1E-04\(\pm\)2E-07 & 3.8E-04\(\pm\)5E-07 \\
 & $p_2$ & 6.6E-02\(\pm\)3E-05 & 5.7E-02\(\pm\)6E-05 & 7.0E-02\(\pm\)6E-05 & 5.5E-02\(\pm\)4E-05 & 5.7E-02\(\pm\)3E-05 \\
 & $p_3$ & 2.9E-05\(\pm\)1E-08 & 2.8E-05\(\pm\)4E-08 & 2.3E-05\(\pm\)4E-08 & 2.6E-05\(\pm\)5E-08 & 3.3E-05\(\pm\)2E-08 \\
 & $p_4$ & 6.1E-02\(\pm\)5E-06 & 6.0E-02\(\pm\)1E-05 & 6.2E-02\(\pm\)1E-05 & 6.0E-02\(\pm\)2E-05 & 5.9E-02\(\pm\)6E-06 \\
 & $p_5$ & 9.4E-02\(\pm\)7E-05 & 9.1E-02\(\pm\)9E-05 & 9.7E-02\(\pm\)7E-05 & 9.9E-02\(\pm\)4E-04 & 9.6E-02\(\pm\)5E-05 \\
 & $p_6$ & 1.1E-02\(\pm\)5E-06 & 9.7E-03\(\pm\)2E-05 & 9.1E-03\(\pm\)1E-05 & 8.6E-03\(\pm\)3E-05 & 1.0E-02\(\pm\)6E-06 \\
 & $p_7$ & 1.4E-03\(\pm\)1E-06 & 1.6E-03\(\pm\)1E-06 & 1.5E-03\(\pm\)1E-06 & 1.5E-03\(\pm\)5E-06 & 1.5E-03\(\pm\)1E-06 \\
 & $p_8$ & 2.5E-02\(\pm\)8E-06 & 2.5E-02\(\pm\)1E-05 & 2.4E-02\(\pm\)5E-06 & 2.6E-02\(\pm\)3E-05 & 2.5E-02\(\pm\)7E-06 \\
\end{tabular}

     \caption{The mean and standard deviation of the least-squares parameter estimates (Equation~\ref{eqn:parameter_estimation}) used in Case I, where the maximal conductance, is misspecified by scaling it with \(\lambda\). These were obtained from each training protocol (\(d_1\)--\(d_5\)) for multiple repeats of synthetically generated data.}
     \label{tab:CaseI_parameters}
 \end{table}

 \begin{table}[]
     \centering
     \begin{tabular}{lccc}
\toprule
$\lambda$ & Mean interval width (nA) & DGP in interval (\%) & Midpoint RMSE (nA) \\\midrule
0.25 & 7.4E-02 \(\pm\) 1.1E-04 & 3.4E01 \(\pm\) 2.3E-02 & 1.6E-01 \(\pm\) 8.9E-05 \\
0.30 & 6.6E-02 \(\pm\) 6.5E-05 & 3.7E01 \(\pm\) 5.0E-02 & 1.4E-01 \(\pm\) 8.9E-05 \\
0.35 & 6.1E-02 \(\pm\) 5.8E-05 & 4.2E01 \(\pm\) 4.3E-02 & 1.3E-01 \(\pm\) 9.1E-05 \\
0.42 & 5.6E-02 \(\pm\) 6.4E-05 & 5.0E01 \(\pm\)4.2E-02 & 1.1E-01  \(\pm\) 9.2E-05 \\
0.50 & 4.9E-02  \(\pm\)7.1E-05 & 5.1E01  \(\pm\)5.1E-02 & 8.5E-02 \(\pm\) 9.4E-05 \\
0.59 & 4.0E-02  \(\pm\) 8.4E-05 & 5.3E01\(\pm\) 6.7E-02 & 6.5E-02 \(\pm\) 9.7E-05 \\
0.71 & 2.8E-02 \(\pm\) 9.1E-05 & 5.5E01 \(\pm\) 8.8E-02 & 4.7E-02 \(\pm\) 9.0E-05 \\
0.84 & 1.4E-02 \(\pm\) 9.3E-05 & 5.5E01 \(\pm\) 1.7E-01 & 3.4E-02 \(\pm\) 8.4E-05 \\
1.00 & 2.1E-04 \(\pm\) 3.7E-05 & 9.4E01 \(\pm\) 9.2E-00 & 3.0E-02 \(\pm\) 6.7E-05 \\
1.19 & 1.2E-02 \(\pm\) 8.9E-05 & 5.6E01 \(\pm\) 6.9E-01 & 3.3E-02 \(\pm\) 5.1E-05 \\
1.41 & 2.1E-02 \(\pm\) 8.5E-05 & 5.7E01 \(\pm\) 6.7E-01 & 3.7E-02 \(\pm\) 5.3E-05 \\
1.68 & 2.8E-02 \(\pm\) 8.5E-05 & 5.7E01 \(\pm\) 1.7E-00 & 4.2E-02 \(\pm\) 6.2E-05 \\
2.00 & 3.3E-02 \(\pm\) 8.6E-05 & 5.6E01 \(\pm\) 1.7E-00 & 4.7E-02 \(\pm\) 6.9E-05 \\
2.38 & 3.6E-02 \(\pm\) 8.6E-05 & 5.5E01 \(\pm\) 1.4E-00 & 5.1E-02 \(\pm\) 7.3E-05 \\
2.83 & 3.9E-02 \(\pm\) 8.4E-05 & 5.4E01  \(\pm\)1.4E-00 & 5.4E-02 \(\pm\) 7.5E-05 \\
3.36 & 4.1E-02 \(\pm\) 8.6E-05 & 5.4E01 \(\pm\) 1.7E-01 & 5.7E-02 \(\pm\) 7.6E-05 \\
4.00 & 4.3E-02 \(\pm\) 8.4E-05 & 5.3E01 \(\pm\) 1.5E-01 & 5.9E-02 \(\pm\) 7.8E-05 \\
\bottomrule
\end{tabular}

     \caption{A summary of showing how the spread-of-predictions interval (Equation~\ref{eqn:spread_of_prediction_interval}) behaves under Case I. Here we show: the mean width of the interval (averaged over each observation time); the proportion of observations for which the the underlying DGP (minus noise) lies within the interval; the RMSE  between the data and the midpoint prediction (Equation~\ref{eqn:midpoint_prediction}). By considering ten randomly sampled datasets (each containing a repeat each protocol \(d_1\)--\(d_5\)), we show the mean and standard deviation of these values. }
     \label{tab:CaseI_summary}
 \end{table}

\subsection{Appendix C: Further Case II results}  \label{sec:appendix_C}
Table~\ref{tab:CaseII_Beattie_parameters} and Table~\ref{tab:CaseII_Wang_parameters} summarise the parameter estimates obtained in Section~\ref{sec:caseII} using the Beattie model and Wang model, respectively. Here, the Wang model was chosen as the DGP and so, the Wang model is an example of a correctly specified model, whereas the Beattie model is a discrepant model. This is reflected by the parameter estimates which show that when the Wang model is fitted to the data, we obtain similar parameter estimates from each protocol, whereas our parameter estimates for the Beattie model are dependent on the protocol used for training. 
 \begin{table}[]
     \centering    
      \setlength{\tabcolsep}{5pt}
      \begin{tabular}{lccccc}
\toprule
 & $d_1$ & $d_2$ & $d_3$ & $d_4$ & $d_5$ \\ \midrule
$g$ & 1.5E-01\(\pm\)3E-05 & 1.5E-01\(\pm\)5E-05 & 1.6E-01\(\pm\)3E-05 & 1.5E-01\(\pm\)5E-05 & 1.6E-01\(\pm\)2E-05 \\
$p_1$ & 1.6E-03\(\pm\)7E-07 & 1.6E-03\(\pm\)8E-07 & 1.7E-03\(\pm\)1E-06 & 1.7E-03\(\pm\)1E-06 & 2.0E-03\(\pm\)5E-07 \\
$p_2$ & 7.3E-02\(\pm\)2E-05 & 7.9E-02\(\pm\)2E-05 & 3.7E-02\(\pm\)2E-05 & 8.4E-02\(\pm\)4E-05 & 5.4E-02\(\pm\)2E-05 \\
$p_3$ & 1.9E-05\(\pm\)2E-08 & 2.2E-05\(\pm\)2E-08 & 4.3E-05\(\pm\)6E-08 & 2.2E-05\(\pm\)2E-08 & 3.0E-05\(\pm\)2E-08 \\
$p_4$ & 5.2E-02\(\pm\)7E-06 & 5.1E-02\(\pm\)1E-05 & 4.6E-02\(\pm\)1E-05 & 5.1E-02\(\pm\)1E-05 & 4.9E-02\(\pm\)6E-06 \\
$p_5$ & 1.1E-01\(\pm\)7E-05 & 9.6E-02\(\pm\)2E-05 & 9.3E-02\(\pm\)2E-05 & 1.2E-01\(\pm\)2E-04 & 9.6E-02\(\pm\)6E-05 \\
$p_6$ & 2.3E-02\(\pm\)5E-06 & 2.4E-02\(\pm\)8E-06 & 2.2E-02\(\pm\)5E-06 & 2.7E-02\(\pm\)2E-05 & 2.3E-02\(\pm\)7E-06 \\
$p_7$ & 8.9E-03\(\pm\)5E-06 & 7.1E-03\(\pm\)3E-06 & 7.4E-03\(\pm\)2E-06 & 8.9E-03\(\pm\)1E-05 & 6.8E-03\(\pm\)5E-06 \\
$p_8$ & 2.9E-02\(\pm\)8E-06 & 3.1E-02\(\pm\)1E-05 & 3.0E-02\(\pm\)7E-06 & 2.9E-02\(\pm\)2E-05 & 3.1E-02\(\pm\)7E-06 \\
\bottomrule
\end{tabular}
      \caption{The parameter estimates obtained for Case II (Section~\ref{sec:CaseI}) when using the Beattie model to fit data generated by the Wang model.}
     \label{tab:CaseII_Beattie_parameters}
 \end{table}

 \begin{table}[]
     \centering    
     \setlength{\tabcolsep}{5pt}
      \begin{tabular}{l c c c c c}
\toprule
 & $d_1$ & $d_2$ & $d_3$ & $d_4$ & $d_5$ \\
 \midrule
$g$ & 1.5E-01\(\pm\)3E-05 & 1.5E-01\(\pm\)4E-05 & 1.5E-01\(\pm\)2E-05 & 1.5E-01\(\pm\)4E-05 & 1.5E-01\(\pm\)2E-05 \\
$k_b$ & 3.7E-02\(\pm\)4E-04 & 3.6E-02\(\pm\)1E-03 & 3.7E-02\(\pm\)2E-04 & 3.6E-02\(\pm\)2E-03 & 3.7E-02\(\pm\)3E-04 \\
$k_f$ & 2.4E-02\(\pm\)9E-05 & 2.4E-02\(\pm\)4E-04 & 2.4E-02\(\pm\)9E-05 & 2.4E-02\(\pm\)6E-04 & 2.4E-02\(\pm\)7E-05 \\
$q_1$ & 9.1E-02\(\pm\)5E-05 & 9.1E-02\(\pm\)7E-05 & 9.1E-02\(\pm\)2E-05 & 9.1E-02\(\pm\)1E-04 & 9.1E-02\(\pm\)6E-05 \\
$q_2$ & 2.3E-02\(\pm\)6E-06 & 2.3E-02\(\pm\)1E-05 & 2.3E-02\(\pm\)5E-06 & 2.3E-02\(\pm\)1E-05 & 2.3E-02\(\pm\)8E-06 \\
$q_3$ & 2.2E-02\(\pm\)5E-04 & 2.3E-02\(\pm\)7E-04 & 2.2E-02\(\pm\)3E-04 & 2.3E-02\(\pm\)8E-04 & 2.2E-02\(\pm\)3E-04 \\
$q_4$ & 1.2E-02\(\pm\)4E-04 & 1.1E-02\(\pm\)7E-04 & 1.2E-02\(\pm\)2E-04 & 1.1E-02\(\pm\)8E-04 & 1.2E-02\(\pm\)2E-04 \\
$q_5$ & 1.4E-02\(\pm\)2E-04 & 1.4E-02\(\pm\)4E-04 & 1.4E-02\(\pm\)1E-04 & 1.4E-02\(\pm\)7E-04 & 1.4E-02\(\pm\)1E-04 \\
$q_6$ & 3.8E-02\(\pm\)2E-04 & 3.8E-02\(\pm\)3E-04 & 3.8E-02\(\pm\)1E-04 & 3.8E-02\(\pm\)6E-04 & 3.8E-02\(\pm\)2E-04 \\
$q_7$ & 6.9E-05\(\pm\)7E-08 & 6.9E-05\(\pm\)3E-07 & 6.9E-05\(\pm\)1E-07 & 6.9E-05\(\pm\)4E-07 & 6.9E-05\(\pm\)1E-07 \\
$q_8$ & 4.2E-02\(\pm\)8E-06 & 4.2E-02\(\pm\)4E-05 & 4.2E-02\(\pm\)1E-05 & 4.2E-02\(\pm\)5E-05 & 4.2E-02\(\pm\)1E-05 \\
$q_9$ & 6.5E-03\(\pm\)4E-06 & 6.5E-03\(\pm\)5E-06 & 6.5E-03\(\pm\)2E-06 & 6.5E-03\(\pm\)8E-06 & 6.5E-03\(\pm\)7E-06 \\
$q_{10}$ & 3.3E-02\(\pm\)7E-06 & 3.3E-02\(\pm\)2E-05 & 3.3E-02\(\pm\)9E-06 & 3.3E-02\(\pm\)2E-05 & 3.3E-02\(\pm\)8E-06 \\
$q_{11}$ & 4.7E-02\(\pm\)1E-03 & 4.9E-02\(\pm\)3E-03 & 4.7E-02\(\pm\)4E-04 & 4.9E-02\(\pm\)3E-03 & 4.7E-02\(\pm\)6E-04 \\
$q_{12}$ & 6.3E-02\(\pm\)4E-04 & 6.3E-02\(\pm\)6E-04 & 6.3E-02\(\pm\)4E-04 & 6.3E-02\(\pm\)6E-04 & 6.3E-02\(\pm\)2E-04 \\
\bottomrule
\end{tabular}
      \caption{The parameter estimates obtained for Case II when using the correct Wang model to fit its synthetic data.}
     \label{tab:CaseII_Wang_parameters}
 \end{table}

Table~\ref{tab:CaseII_summary} shows the behaviour of our spread-of-prediction intervals (Equation~\ref{eqn:spread_of_prediction_interval}) for both the Wang model and the Beattie model as described in Section~\ref{sec:caseII}. Here, we see that the average width of this interval (averaged over the length of the protocol) is much larger for the Beattie model (a discrepant model) when compared with the Wang model (the same model used for data generation).
  \begin{table}[]
     \centering    
      \begin{tabular}{llll}
\toprule
Model & Mean interval width (nA) & DGP in interval (\%) & Midpoint RMSE (nA) \\ \midrule
Beattie & 7.5E-02\(\pm\)9E-05 & 3.4E01\(\pm\)7E-02 & 1.1E-01\(\pm\)8E-05 \\
Wang & 7.0E-04\(\pm\)2E-04 & 8.7E01\(\pm\)2E01 & 3.0E-02\(\pm\)2E-05 \\
\bottomrule
\end{tabular}
      \caption{A summary showing how the spread-of-predictions interval (Equation~\ref{eqn:spread_of_prediction_interval}) behaves under Case II (Section~\ref{sec:caseII}), for both the Beattie model (a discrepant model) and the Wang model (a correctly specified model). The columns show: the mean width of the interval (averaged over each observation time); the proportion of observations for which the the underlying DGP (minus noise) lies within the interval; the RMSE  between the data and the midpoint prediction (Equation~\ref{eqn:midpoint_prediction}). By considering ten randomly sampled datasets (each containing a repeat each protocol \(d_1\)--\(d_5\)), we show the mean and standard deviation of these values.}
     \label{tab:CaseII_summary}
 \end{table}

\end{document}